\documentclass[preprint,12pt]{elsarticle}
\usepackage{amssymb}
\usepackage{amsmath}
\usepackage{multirow}
\usepackage{url}

\usepackage[defaultcolor=red]{changes}  
\definechangesauthor[name={li}, color=red]{A} 

\journal{Computer Methods and Programs in Biomedicine}

\begin{document}

\begin{frontmatter}

\title{Simulation-to-Real First-Break Segmentation for Efficient Inversion in Musculoskeletal Ultrasound Tomography
}

\author[label1,label2,label3,label4]{Yifei Sun}
\author[label1,label2,label3]{Yubing Li\corref{cor1}}
\ead{liyubing@mail.ioa.ac.cn}
\author[label4]{Yannick Benezeth}
\author[label4]{Stéphanie Bricq}
\author[label1,label2,label3]{Yunrong Zhang}
\author[label1,label2,label3]{Lekang Jiang}
\author[label1,label2,label3]{Chang Su}
\author[label5]{Ligang Cui}
\author[label1,label2,label3]{Weijun Lin\corref{cor1}}
\ead{linwj@mail.ioa.ac.cn}

\cortext[cor1]{Corresponding authors.}

\affiliation[label1]{
organization={State Key Laboratory of Acoustics and Marine Information, Institute of Acoustics, Chinese Academy of Sciences},
city={Beijing},
postcode={100190},
country={China}
}

\affiliation[label2]{
organization={Laboratory of Ultrasonics, Institute of Acoustics, Chinese Academy of Sciences},
city={Beijing},
postcode={100190},
country={China}
}

\affiliation[label3]{
organization={University of Chinese Academy of Sciences},
city={Beijing},
postcode={100049},
country={China}
}

\affiliation[label4]{
organization={Université Bourgogne Europe, IMVIA UR 7535},
city={Dijon},
postcode={21000},
country={France}
}

\affiliation[label5]{
organization={Department of Ultrasound, Peking University Third Hospital},
city={Beijing},
postcode={100191},
country={China}
}
\begin{abstract}
Full-waveform inversion (FWI) has become a promising strategy for quantitative musculoskeletal ultrasound computed tomography (USCT), a safe and accessible imaging modality. However, in anatomically complex regions, bone-related scattering, attenuation, and signal degradation make FWI highly sensitive to the accuracy of the initial acoustic-property distributions: poor initialization can lead to cycle skipping. First-arrival traveltimes or times of flight provide important kinematic information for initial-model construction, yet conventional trace-wise picking often fails because these signals are weak, spatially heterogeneous, and frequently buried in system noise. We propose a learning-assisted reconstruction pipeline that couples segmentation-based first-arrival extraction with hybrid full-waveform inversion (HFWI), which combines waveform fitting with Rytov-approximation-based traveltime information during the early inversion stage. By treating the first-arrival trajectory across receiver channels as a two-dimensional first-break segmentation target, a lightweight U-Net exploits its spatial continuity rather than process each receiver trace independently. To address both the scarcity of manually annotated experimental data and the simulation-to-real gap, the network is pretrained entirely on task-specific simulations augmented with real system-noise recordings. A stage-wise training strategy progressively introduces realistic signal degradation, followed by decoder-only fine-tuning using limited weakly labeled experimental data. The method is evaluated on \textit{in vitro} phantom, \textit{ex vivo} bovine-limb, and \textit{in vivo} human-thigh datasets. Compared with conventional short-term average/long-term average picking, the proposed network produces more spatially coherent first-arrival trajectories, reduces the mean extraction error, and processes a full-matrix-capture dataset within seconds. When integrated into HFWI, the extracted first-arrival travetimes improve initial-model construction and lead to more stable subsequent FWI reconstructions, even for challenging cases in which the estimated local first-arrival SNRs are below 3 dB.
\end{abstract}

\begin{highlights}
\item First-arrival extraction is formulated as a segmentation task that exploits spatial continuity across receiver channels.
\item Stage-wise synthetic pretraining and decoder-only fine-tuning enable simulation-to-real adaptation using limited weakly labeled experimental data.
\item Network-derived first arrivals improve full waveform inversion initialization and reconstruction stability.
\end{highlights}

\begin{keyword}
Musculoskeletal ultrasound computed tomography \sep First-arrival extraction \sep Full-waveform inversion
\end{keyword}

\end{frontmatter}



\section{Introduction}
Musculoskeletal soft-tissue disorders are a major cause of pain and functional impairment and often require high-resolution imaging for accurate diagnosis and treatment assessment \cite{ref1}. Ultrasound computed tomography (USCT) has recently emerged as a candidate technique for such musculoskeletal evaluations \cite{ref2,ref3}. By reconstructing quantitative acoustic-property distributions, particularly sound speed and attenuation, from transmit--receive array measurements \cite{ref4}, USCT effectively visualizes pathophysiological conditions in muscle, tendon, and fascia \cite{ref5,ref6}. Compared with X-ray computed tomography (X-CT) and magnetic resonance imaging (MRI), ultrasound is free of ionizing radiation, compatible with metallic implants, and comparatively cost effective. These attributes underpin its potential in sports medicine scenarios, where ultrasound is already widely used for assessing musculoskeletal soft-tissue, joint, and peripheral-nerve assessment \cite{ref7,ref8}.

USCT has demonstrated high-resolution quantitative imaging performance in breast and other predominantly soft-tissue applications \cite{ref9,ref10,ref11}. Its extension to musculoskeletal imaging, however, is substantially more challenging because bone introduces strong acoustic impedance contrast, high sound speed, attenuation, and complex scattering. These effects distort transmitted waveforms and can severely reduce the signal-to-noise ratio (SNR) along propagation paths traversing or interacting with bone. Previous efforts have demonstrated the feasibility of ultrasound imaging and tomography in the presence of bone-related structures \cite{ref2,ref3,ref12}; however, a robust USCT framework for quantitative assessment in the musculoskeletal system remains under development.

Full-waveform inversion (FWI) has increasingly been investigated for this purpose \cite{ref3,ref13}. FWI iteratively updates a given acoustic model by minimizing the discrepancy between measured and simulated data. Because the forward kernel accounts for wave phenomena such as diffraction and multiple scattering within the assumed physical description \cite{ref14,ref15}, FWI provides a promising framework for imaging strongly heterogeneous media. With sufficiently informative data, its spatial resolution can approach the subwavelength scale \cite{ref15}. In practice, dense transmit--receiver sampling is typically required, motivating the use of ring- or cylinder-shaped transducer arrays operated in full-matrix-capture (FMC) mode, in which each element transmits sequentially while all channels receive.

A major limitation of FWI is the strong non-convexity of waveform matching. When the phase discrepancy between measured and simulated data becomes sufficiently large, the inversion can converge toward an incorrect local minimum, a phenomenon commonly referred to as cycle skipping \cite{ref15,ref16}. This issue is particularly severe in musculoskeletal USCT because the large acoustic contrast between bone and surrounding soft tissues can produce substantial phase errors when the starting model differs from the true sound-speed distribution. Several strategies have been proposed to reduce this sensitivity, including the use of lower frequencies \cite{ref3,ref15}, improved initial models \cite{ref17,ref18}, and alternative misfit functions designed to reduce phase sensitivity or reshape the optimization landscape \cite{ref15,Misfit1}. Because extension toward lower frequencies is ultimately constrained by transducer bandwidth and acquisition hardware, improving the early-stage kinematic information through algorithmic methods provides an attractive complementary route.

First-arrival traveltimes are particularly useful in this context because they contain predominantly kinematic information and can constrain the large-scale sound-speed distribution before or during waveform inversion \cite{ref10,ref17}. The onset of the first-arriving wave at each receiver, commonly referred to as the first break, provides an estimate of the corresponding first-arrival time. These measurements can be used in Eikonal-based traveltime tomography or incorporated directly into a hybrid inversion objective to guide the recovery of the low-wavenumber background model. Reliable first-break picking is therefore important for reducing the initial phase mismatch and improving subsequent FWI convergence.

Automatic first-break extraction, however, is difficult in musculoskeletal USCT. Bone-related attenuation and scattering can make the earliest arrivals weak, spatially heterogeneous, and locally discontinuous across receiver channels, while system noise and later-arriving waveforms can obscure their onset. Conventional model-driven methods, including the Akaike Information Criterion (AIC) \cite{ref19,ref20,ref21}, short-term average/long-term average (STA/LTA) schemes \cite{ref22,ref23,ref24}, and cross-correlation-based methods \cite{ref25,ref26}, rely on handcrafted characteristic functions, thresholds, or local waveform assumptions. Their performance can consequently deteriorate under low-SNR conditions and often depends on case-specific parameter tuning.

Differently, data-driven or learning-based methods provide an alternative by learning discriminative features directly from data \cite{ref27,ref28}. Existing approaches generally formulate first-break extraction as either regression or segmentation. Regression-based methods estimate an arrival time from individual traces or shot gathers \cite{ref29,ref30}, whereas segmentation-based methods treat a collection of signals from neighboring traces as a structured input and identify the first-break trajectory in a two-dimensional (2-D) representation \cite{ref31,ref32,ref33,ref34}. The latter formulation is particularly attractive for FMC USCT because first breaks from adjacent receivers exhibit strong spatial continuity even when individual traces have poor SNR. Exploiting this continuity can make the extraction less dependent on the local waveform amplitude of a single channel. Learning-based methods can reduce manual intervention during inference and have shown improved performance over conventional model-driven techniques in several studies\cite{ref30,ref33,ref34}. Nevertheless, learning-based methods introduce a different bottleneck: robust performance typically requires large amounts of representative labeled data. Acquiring such experimental datasets and manually annotating first breaks are time consuming, costly, and prone to inter-observer labeling inconsistency.

Simulation provides a natural way to alleviate this limitation. Numerical wave propagation can generate large quantities of task-specific data with accurately known first-break labels, avoiding exhaustive experimental annotation. The difficulty is that a network trained only on simulated data must ultimately operate on measurements acquired from a physical system. Differences or simplifications in anatomical variability, wave-propagation physics, transducer responses, electronics, noise, and other acquisition effects lead to a discrepancy between synthetic and experimental data distributions, commonly referred to as the simulation-to-real (Sim2Real) gap. Transfer learning can partially address such domain shifts \cite{ref35,ref36,ref37}. However, its success depends on the similarity between source and target domains, and the first-break extraction performance can degrade when this match is poor \cite{ref38,ref39}. This is particularly relevant to USCT, for which large, representative experimental datasets are scarce, acquisition-system dependent, and difficult to share. Recent studies have shown that networks trained predominantly on simulated radio-frequency data can be transferred to experimental USCT measurements \cite{refSim2Real1,refSim2Real2}, suggesting that physics-based simulation can serve as an effective training source when the Sim2Real gap is explicitly considered.

In this study, we develop a simulation-to-real first-break segmentation framework for musculoskeletal USCT and integrate it with previously developed hybrid full-waveform inversion (HFWI) \cite{ref41}. Instead of detecting the first break independently on each receiver trace, we formulate the first-break trajectory across receiver channels as a 2-D segmentation problem and use a lightweight U-Net to exploit its spatial continuity. The network is pretrained entirely on task-specific synthetic data generated from anatomical numerical phantoms with augmentation using measured system noises. To reduce the simulation-to-real discrepancy, the network is subsequently fine-tuned using a limited set of weakly labeled experimental data. The resulting first-arrival times are then incorporated into HFWI to provide kinematic guidance during the early inversion stage. The framework is evaluated using \textit{in vitro}, \textit{ex vivo}, and \textit{in vivo} cases, assessing both first-break extraction and its downstream effect on inversion. Compared with conventional STA/LTA picking, the proposed method lead to faster convergence and higher-quality in FWI reconstructions. The results demonstrate a practical simulation-to-real route for obtaining reliable first-break information in musculoskeletal USCT without requiring large manually annotated experimental datasets.

The remainder of this article is organized as follows. Section~2 presents the overall imaging framework, including first-break segmentation and subsequent HFWI method. In Section~3, we introduce the dataset composition and training strategy, and evaluate the picking accuracy and reconstruction performance using \textit{in vitro}, \textit{ex vivo}, and \textit{in vivo} experiments. Section~4 discusses the rationale behind the pipeline design and the observed performance differences. Finally. Section~5 concludes the study.

\section{Methods and Workflow}
\label{sec2}
\subsection{Scanning Setup and Data Acquisition}
\label{sub1sec2}
The musculoskeletal USCT platform combines a self-developed acquisition module (Fig.~1(a)) with a 512-element ring array of 22~cm diameter (Fig.~1(b)). The specimen is centered within the ring array and fully immersed in degassed water for acoustic coupling, as illustrated by the bovine-limb cross-section in Fig.~1(b). Each element operates in both transmit and receive modes, enabling full-matrix recording of transmit--receive channel waveforms. Under 0.9-MHz excitation, the transmitted pulse exhibited a $-6$~dB bandwidth of approximately 0.6--1.1~MHz, as estimated from target-free measurements. Other hardware details are provided in \cite{ref41}.

Each element is 2~cm tall along the axis perpendicular to the imaging plane, favoring in-plane propagation and supporting a 2-D approximation. In \textit{ex vivo} and \textit{in vivo} acquisitions, however, out-of-plane propagation and scattering can reduce the effective signal-to-noise ratio (SNR), particularly in the early arrivals that carry first-arrival information. This motivates the subsequent first-arrival extraction strategy described below.

\begin{figure}[htbp]
\centering
\begin{minipage}[t]{0.48\linewidth}
\centering
\includegraphics[width=\linewidth]{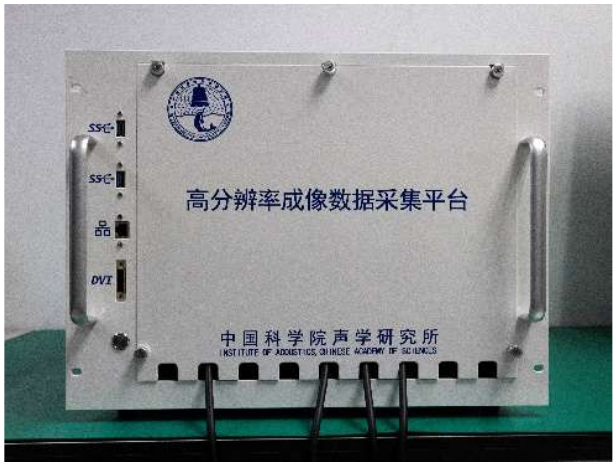}\
\small (a)
\end{minipage}
\hfill
\begin{minipage}[t]{0.48\linewidth}
\centering
\includegraphics[width=\linewidth]{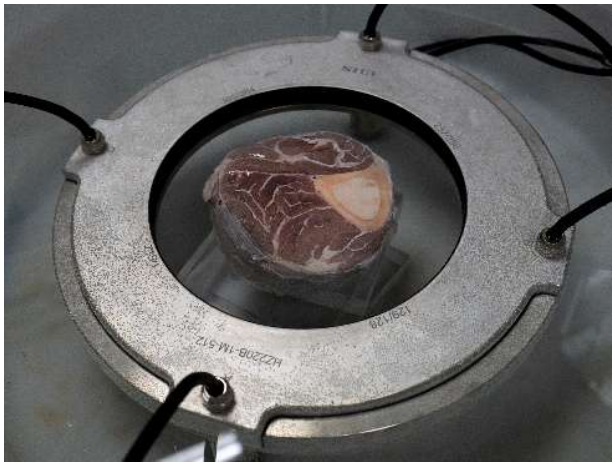}\
\small (b)
\end{minipage}
\caption{Musculoskeletal USCT system. (a) Self-developed 256-channel acquisition module using time-division multiplexing to address 512 array elements. (b) Ring array with 512 transmit/receive elements and a diameter of 22~cm, immersed in water for acoustic coupling.}
\label{fig1}
\end{figure}

\subsection{Overview of First-Arrival-Guided Imaging Pipeline}
\label{sub2sec2}
Assuming an appropriately trained segmentation network, each FMC dataset is processed through three stages to obtain the final sound-speed reconstruction. First, the segmentation network extracts first-arrival times from the recorded channel data (Sec.~\ref{sub3sec2}). Second, during the early stage of HFWI, the extracted first-arrival information is used jointly with the low-frequency components of the measured data, which generally exhibit low SNR, to construct a sound-speed model suitable for subsequent higher-frequency FWI (Sec.~\ref{sub4sec2}). Third, starting from this model, the reconstruction is further refined using higher-frequency components of the measured data without the first-arrival term.

\subsection{Segmentation-Based First-Arrival Extraction}
\label{sub3sec2}

Musculoskeletal USCT measurements can include propagation paths for which the estimated local first-arrival SNR falls below $3$~dB. To improve extraction robustness under such challenging conditions, we formulate first-arrival picking as a segmentation problem. Unlike conventional trace-wise methods that estimate the arrival time independently for each receiver, the segmentation-based formulation exploits the spatial continuity of first arrivals across neighboring channels together with the broader waveform context of each transmit event. The remainder of this subsection describes the network architecture, training protocol, and inference procedure used for the proposed segmentation-based extractor.

\subsubsection{Task Definition and Input/Output Representation}
\label{sub1sub3sec2}
\begin{figure*}[t]
    \centering
    \includegraphics[width=\textwidth]{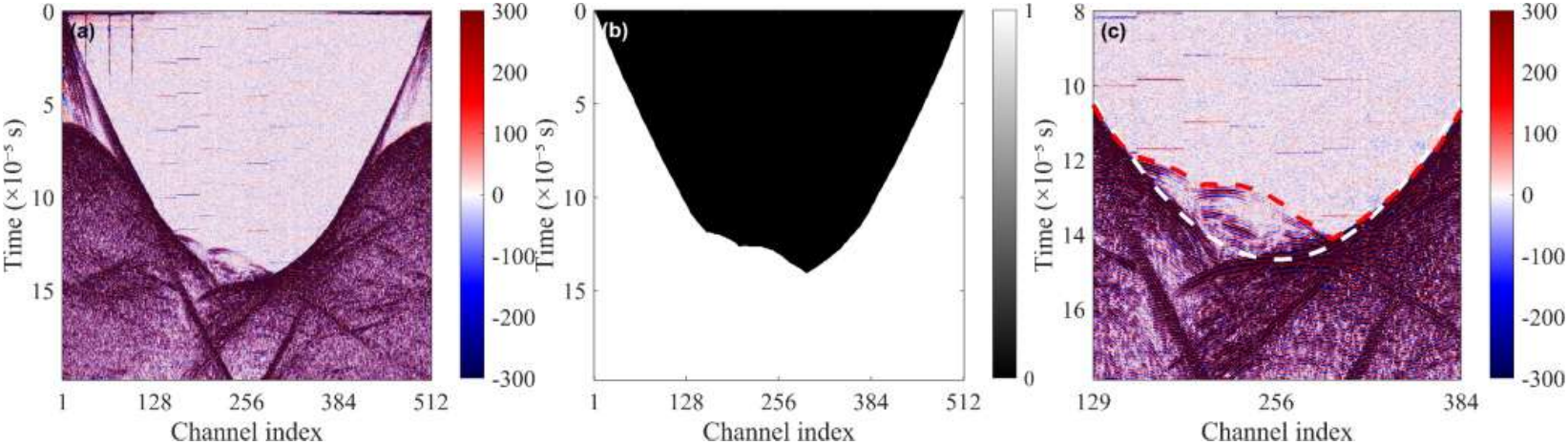}
    \caption{Definition of the segmentation-based first-arrival extraction task. (a) Recorded FMC waveforms for a single transmitting element. (b) Binary segmentation mask, where label~0 denotes the pre-arrival background and label~1 denotes the post-arrival region. (c) Zoomed view showing the extracted first-arrival trajectory (red dashed curve) and the reference trajectory obtained from a water-only scan (white dashed curve).}
    \label{fig2}
\end{figure*}

As illustrated in Fig.~\ref{fig2}(a), the recorded waveforms for a single transmitting element (element~\#1 in this example) are arranged according to receiver index to form a two-dimensional representation with time along the vertical axis and receiver channel along the horizontal axis. To visualize the weak low-SNR components around the first arrivals, a diverging colormap is used with the display dynamic range clipped to $\pm 300$~a.u.

In the proposed binary segmentation formulation, first-break extraction is treated as the separation of the pre-arrival background from the post-arrival region. The network outputs a binary mask, as shown in Fig.~2(b). In the reference mask, label~0 denotes the pre-arrival background, dominated by system noise, whereas label~1 denotes the post-arrival region, extending from the onset of the first arrival to the end of each trace and containing the main acoustic responses. For each receiver channel, the first-arrival time is obtained from the earliest transition from label~0 to label~1 along the time axis. The resulting channel-wise arrival times form the first-arrival trajectory. Once the mask is obtained, the trajectory is recovered and superimposed on the original waveforms as a red dashed curve, as shown in Fig.~2(c). For reference, the white dashed curve shows the trajectory measured from a water-only scan without scatterers. The channel-wise time difference between the two curves represents the delay introduced by the heterogeneous medium, and the area between the two trajectories is referred to as the first-arrival region throughout this paper.

\subsubsection{Network Architecture}
\label{sub2sub3sec2}
The segmentation network is implemented as a lightweight U-Net with three downsampling and three upsampling stages, as illustrated in Fig.~3. The network follows a symmetric encoder--decoder structure: the encoder progressively extracts multiscale features, whereas the decoder restores spatial resolution to generate the segmentation output.

The encoder comprises three downsampling modules. Each module contains two $3\times3$ convolutional layers, each followed by a $2\times2$ max-pooling operation. Each convolutional layer is followed by batch normalization and a rectified linear unit (ReLU) activation. The max-pooling operation reduces the spatial resolution by a factor of two, while the number of feature channels is doubled in the subsequent stage to allow the network to learn increasingly abstract representations.

The decoder mirrors the encoder and comprises three upsampling modules. Each module starts with a transposed convolution layer that doubles the spatial resolution. The corresponding encoder features are then concatenated through a skip connection to preserve fine-scale spatial information. After concatenation, two convolutional layers with the same configuration as those in the encoder are applied. Zero-padding is used in all convolutional layers to maintain consistent spatial dimensions.

To mitigate overfitting, a two-dimensional dropout layer is included after each module, randomly suppressing a fraction of feature channels during training. Finally, a $1\times1$ convolution layer maps the feature representation to class logits. The binary segmentation mask is then obtained from the predicted class label at each pixel, as shown in Fig.~2(b).

\begin{figure}[htbp]
\centering
\includegraphics[width=0.5\linewidth]{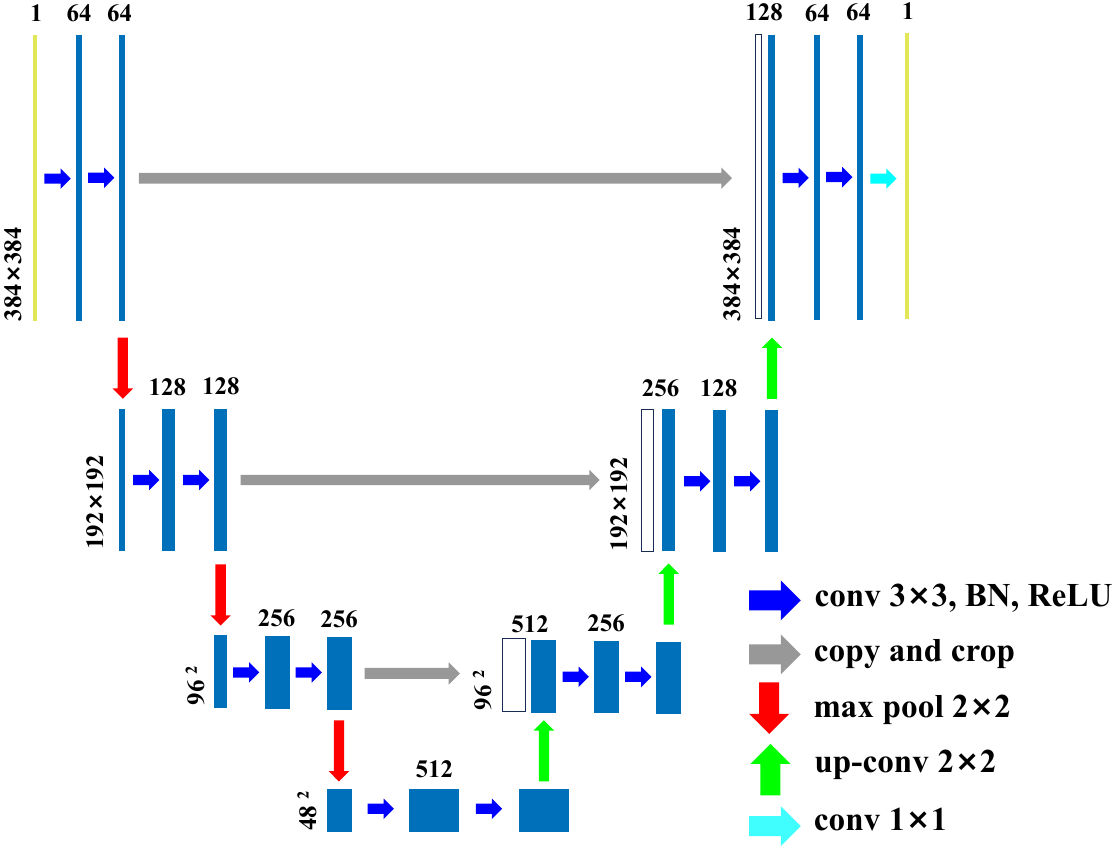}
\caption{Architecture of the lightweight U-Net used for segmentation-based first-arrival extraction.}
\label{fig3}
\end{figure}

\subsubsection{Stage-Wise Pretraining Strategy with Synthetic Data}
\label{sub3sub3sec2}
The segmentation network is ultimately intended for first-break extraction from \textit{in vivo} USCT scans; however, collecting a large, well-labeled clinical dataset is impractical. Synthetic waveforms are therefore used for pretraining. To mitigate the Sim2Real gap, the synthetic data incorporate measured system noise, and a stage-wise pretraining scheme based on curriculum-learning principles progressively increases the difficulty and variability of the training inputs. The following paragraphs describe the synthetic-data generation process and the three pretraining stages.

Each synthetic sample is formed by superimposing measured system noise on simulated multichannel data. The simulated data are generated using numerical solvers and anatomical parameter maps, whereas the noise realization is drawn from idle-mode recordings acquired with the actual USCT system, thereby preserving hardware-specific artifacts and background noise. This construction allows the synthetic samples to reproduce selected characteristics of the experimental measurements while retaining precise control over the first-arrival labels.

The parameter maps used in the simulations, including sound speed and density, are derived from CT Hounsfield units (H) using the following mapping \cite{ref43}:
\begin{equation}
\mu_x = \mu_{\mathrm{lower}} + \frac{H}{1000}
\left(\mu_{\mathrm{upper}}-\mu_{\mathrm{lower}}\right),
\label{eq:hu_mapping}
\end{equation}
where $\mu_x$ marks the mapped acoustic parameter at position $x$, and $\mu_{\mathrm{upper}}$ and $\mu_{\mathrm{lower}}$ represent the corresponding upper and lower bounds of the parameter range. Owing to the limited availability of reliable attenuation measurements in biological tissues, especially in musculoskeletal structures, attenuation is not explicitly modeled in the synthetic simulations. Further details of the digital phantoms are provided in Section~\ref{sub1sec3}.

To enable large-scale training-data generation at manageable computational cost, the synthetic data are generated using a 2-D acoustic wave solver rather than a full 3-D model. To mitigate the resulting discrepancies between synthetic and experimental data, task-specific augmentations are incorporated into a three-stage training strategy that progressively introduces signal characteristics closer to those observed in real measurements.

In the first stage, the network is trained directly on synthetic data to distinguish foreground and background regions based on waveform characteristics.

In the second stage, intensity suppression is applied around the annotated arrival positions to reduce reliance on absolute amplitude. This encourages the network to rely more on wavefront patterns than on absolute intensity, and improves its sensitivity to weak signals around the first-arrival trajectory. The suppression filter is defined as
\begin{equation}
h(t)=s+(1-s)\left[
1-\exp\left(
-\frac{1}{2}
\left(
\frac{t-t_{\mathrm{FA}}}{0.4\gamma L}
\right)^2
\right)
\right],
\label{eq:intensity_suppression}
\end{equation}
where $s\in[0,1]$ controls the minimum residual amplitude, (t) denotes time, and $t_{\mathrm{FA}}$ is the first-arrival time in the current channel. The parameter $L$ corresponds to the time span of the first-arrival region in this channel, defined by the vertical interval between the two dashed curves in Fig.~2(c). To introduce variability in the sharpness of the first-arrival profile, a random scaling factor $\gamma\in[0.4,0.8]$ is applied to $L$ for each channel, thereby modulating the width of the transition region.

In the third stage, domain randomization is introduced within the first-arrival regions to further strengthen the network's use of spatial continuity along the trajectory. Two types of perturbations are randomly added to the simulated waveforms while keeping the segmentation labels unchanged. The first introduces localized noise within a narrow band around the trajectory, whereas the second masks out small segments to create artificial gaps. Since these perturbations do not alter the annotated arrival positions, the original labels remain valid. These perturbations encourage the model to focus on global wavefront patterns rather than overfitting to localized waveform features.

This stage-wise training strategy enables the network to progress from idealized synthetic inputs to more challenging signal conditions. The resulting model provides a pretrained initialization for subsequent fine-tuning with weakly labeled experimental data.

\subsubsection{Fine-Tuning with Experimental Data}
\label{sub4sub3sec2}
Residual mismatches remain after stage-wise pretraining because simulated waveforms cannot fully capture the anatomical complexity and propagation physics of \textit{in vivo} scans. These differences may affect the first-arrival region. To compensate for this residual domain gap, a final fine-tuning step is performed using half of one full FMC dataset collected on the experimental system, corresponding to 256 transmissions and all receive channels. Weak labels are generated by smoothing expert-traced first-arrival trajectories. During fine-tuning, the encoder is frozen to preserve the representations learned during synthetic pretraining, and only the decoder is updated with a reduced learning rate to avoid overfitting. The effect of this fine-tuning step is evaluated in Section~\ref{sec4}.

\subsubsection{Data Processing and Augmentation}
\label{sub5sub3sec2}
To remove the dependence on the absolute element index, the receiver channels are circularly reordered for each transmission so that the transmitting element is always assigned to channel~\#1, followed by the remaining elements in clockwise order. For example, when element~\#300 transmits, the reordered channels correspond sequentially to elements~\#300, \#301, $\ldots$, \#512, \#1, $\ldots$, and \#299.

After this reordering, the 64 receiver elements nearest to the transmitter on each side of the ring are excluded because these channels are not used in the subsequent first-arrival-based inversion. The remaining 384 channels, corresponding to reordered channels~\#65--\#448, are retained as the network input. This procedure provides a consistent transmitter-centered representation and ensures that the segmentation input matches the receiver aperture used in the subsequent inversion.



For first-arrival-based inversion, signals received by approximately 128 neighboring elements around the transmitter are not used in the subsequent reconstruction. Therefore, only the 384 channels approximately opposite to the transmitter, corresponding to reordered channels~\#65 to~\#448, are retained as the input for the segmentation task. This selection focuses the network input on the channel range used for subsequent first-arrival-based inversion.

Along the temporal axis, 384 sampling points are extracted starting from $3.6\times10^{-5}$~s with a sampling interval of $1.2\times10^{-7}$~s, resulting in a $384\times384$ matrix as the network input. Before training, each sample is normalized by its maximum absolute value. This normalization preserves the near-zero baseline of system noise and avoids excessive dynamic-range compression around the first-arrival region.

To further improve generalization, data augmentation is applied during training by flipping each input sample and its corresponding label along the channel dimension. This augmentation increases the diversity of training samples without changing the underlying first-arrival structure.

\subsubsection{Training Settings}
\label{sub6sub3sec2}

All models are implemented in PyTorch and trained on an NVIDIA RTX 5090 GPU. The U-Net architecture uses a base channel width of 64, with the number of filters doubled at each downsampling stage. Each convolutional layer is followed by batch normalization and a ReLU activation, and a two-dimensional dropout layer with a dropout rate of 0.2 is applied after each module to reduce overfitting.

The network is optimized using the Adam algorithm with an initial learning rate of $1\times10^{-4}$. A learning-rate scheduler reduces the learning rate when the validation loss plateaus. The stage-wise pretraining is performed in three successive stages with 20, 40, and 40 epochs, respectively. During Stage~3, each domain-randomization perturbation is applied with a probability of 0.5. The synthetic dataset is split into training and validation sets with a ratio of 9:1, and a batch size of 32 is used throughout training. The segmentation loss is defined as the pixel-wise cross-entropy between the predicted logits and the ground-truth masks.

After the three-stage pretraining on synthetic data, the model is fine-tuned using weakly labeled experimental samples. During this phase, only the decoder is updated, while the encoder remains frozen. Fine-tuning is performed for 50 epochs using a reduced learning rate of $1\times10^{-5}$.

\subsection{Initial Model Construction and Final Imaging via HFWI}
\label{sub4sec2}

After first-arrival times are extracted, they are used to construct an initial sound-speed model for subsequent frequency-domain FWI. In musculoskeletal imaging, however, the large acoustic contrasts among adipose tissue, muscle, and cortical bone pose challenges for conventional Eikonal-based traveltime inversion. Such methods often require careful tuning of regularization terms and may become unstable under highly heterogeneous conditions.

To address this issue, the previously described hybrid full-waveform inversion (HFWI) framework is adopted \cite{ref41}. HFWI incorporates a Rytov-approximation-based traveltime matching term into the conventional FWI objective, allowing a gradual transition from traveltime-dominated fitting to waveform-dominated fitting through a tunable balance parameter. The hybrid loss function is defined as
\begin{equation}
J_{\mathrm{HFWI}}(\mathbf{m},\omega)
=
\sum_s \sum_r
\left[
(1-\alpha)
\left\lVert
d_{\mathrm{cal}}^{s,r}(\mathbf{m},\omega)
-
d_{\mathrm{obs}}^{s,r}
\right\rVert
+
\alpha
\left\lVert
T_{\mathrm{cal}}^{s,r}(\mathbf{m},\omega)
-
T_{\mathrm{obs}}^{s,r}
\right\rVert
\right],
\label{eq:hfwi_loss}
\end{equation}

where $\mathbf{m}$ denotes the model parameter, defined here as slowness. The indices $s$ and $r$ refer to the source and receiver elements, respectively. The terms $d_{\mathrm{cal}}$ and $T_{\mathrm{cal}}$ denote the modeled waveform and the predicted first-arrival time for the current model, whereas $d_{\mathrm{obs}}$ and $T_{\mathrm{obs}}$ denote the observed waveform and the segmentation-derived first-arrival time. The parameter $\alpha\in[0,1]$ controls the trade-off between waveform and traveltime fidelity.

The gradients of the waveform and traveltime terms are computed within the same adjoint-state optimization framework \cite{ref41}. To use the stability of kinematic fitting at low frequencies while retaining the resolution benefit of waveform fitting at higher frequencies, HFWI employs a frequency-dependent weighting strategy. At lower frequencies, where the measured waveforms have lower SNR, a relatively large value of $\alpha$ is used to emphasize the traveltime term. As the frequency increases and the waveform SNR improves, $\alpha$ is progressively reduced. Once sufficiently high-SNR frequencies are reached, $\alpha$ is set to zero, and the reconstruction proceeds as conventional FWI for high-resolution model refinement.

\section{Results}
\label{sec3}
\subsection{Synthetic Dataset Construction and Stage-Wise Input Adaptation}
\label{sub1sec3}
The synthetic dataset used for network training consists of simulated waveforms and additive system noise. To capture realistic noise characteristics, 35,840 noise-only recordings are acquired from the experimental USCT system with transmission disabled and under the same receive-chain settings used for experimental acquisition, including a receiver gain of 21~dB. These recordings capture thermal and electronic noise under static acquisition conditions and are randomly sampled and added to the simulated waveforms during dataset construction.

The anatomical models used for waveform simulation are derived from the open-source OpenWaves database \cite{ref42}, which provides synthetic cross-sectional images generated using a diffusion model. We select 2,100 virtual thigh cross-sections as baseline phantoms. The sound-speed and density maps are assigned from the underlying Hounsfield values using Eq.~\ref{eq:hu_mapping}. The sound-speed bounds are set to $c_{\mathrm{lower}}=1380$~m/s and $c_{\mathrm{upper}}=3700$~m/s. Representative sound-speed maps are shown in Fig.~4, illustrating diverse cortical-bone configurations and muscle morphologies that lead to different first-arrival trajectory patterns in the simulated waveforms.

\begin{figure}[htbp]
\centering
\includegraphics[width=0.5\linewidth]{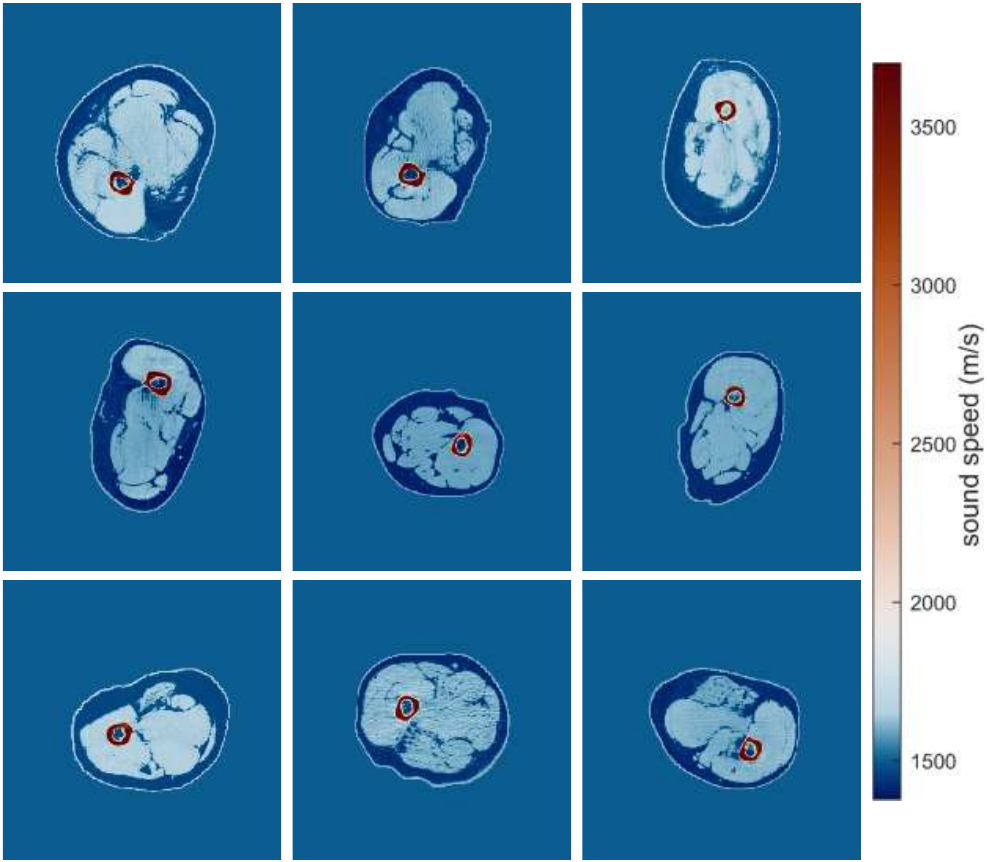}
\caption{Representative sound-speed maps of synthetic thigh phantoms used for network training.}
\label{fig4}
\end{figure}

To widen the range of first-arrival waveform characteristics and emulate inter-subject bone-density differences, sample-wise perturbations are introduced during phantom generation. Specifically, the upper-bound density is sampled from a normal distribution with a mean of 1950~kg/m$^3$ and a standard deviation of 100~kg/m$^3$, whereas the lower-bound density is sampled from a normal distribution with a mean of 900~kg/m$^3$ and a standard deviation of 20~kg/m$^3$. Figure~5 plots the resulting density ranges for 200 randomly selected phantoms.

\begin{figure}[htbp]
\centering
\includegraphics[width=0.45\linewidth]{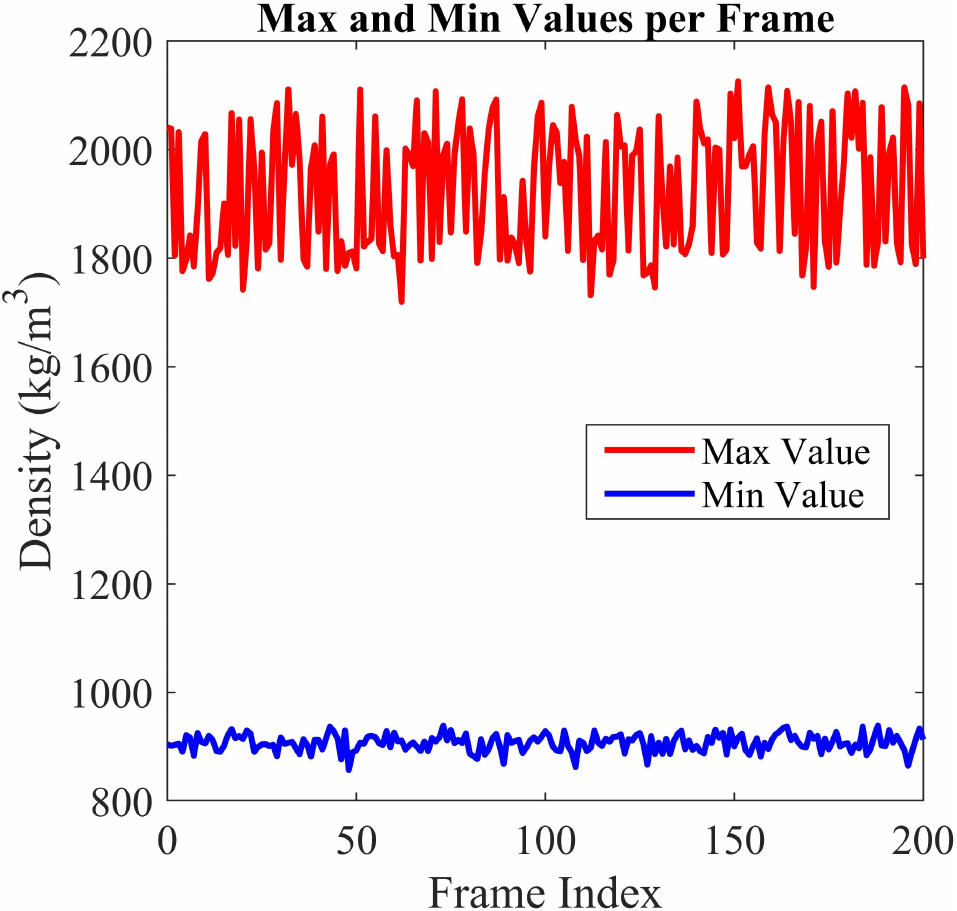}
\caption{Sample-wise density perturbations used in synthetic phantom generation. Maximum and minimum density values are shown for 200 randomly selected phantoms.}
\label{fig5}
\end{figure}

To reduce training redundancy and mitigate overfitting caused by spatially correlated signals from adjacent transmitters, a sparse source-positioning strategy is adopted. For each phantom, 16 transmitting elements are uniformly selected at 32-element intervals around the ring array. For each selected transmitter, a 2-D finite-difference acoustic simulation is performed using the corresponding sound-speed and density maps, and the resulting waveforms recorded at all receiver elements form one two-dimensional input sample. The 2,100 phantoms and 16 source positions yield 33,600 simulated samples. Channel-wise flipping is subsequently applied during training, increasing the effective number of input samples to 67,200.

Before constructing the final dataset, global amplitude normalization is applied to the simulated waveforms to match the intensity range of experimental signals acquired under target-free conditions. This step reduces amplitude-scale mismatch between synthetic and experimental data. For label generation, zero-crossing points of the simulated signals are not used because they can be affected by numerical dispersion. Instead, first-arrival times are computed using the Eikonal equation, yielding consistent annotations across transducer pairs.

Figure~6 illustrates how the synthetic inputs evolve during stage-wise pretraining. Figure~6(a) shows a normalized Stage~1 sample without additional input perturbation. Figures~6(b) and 6(c) show the corresponding inputs from Stage~2 and Stage~3, where intensity suppression and domain randomization are successively introduced. For visual comparison, all images are rendered with the same amplitude window from $-0.02$ to $0.02$. These stage-wise modifications alter the local amplitude and continuity around the first-arrival region while preserving the underlying trajectory, producing synthetic inputs that better approximate the weak and partially discontinuous first-arrival patterns observed in experimental measurements.

\begin{figure}[htbp]
\centering
\includegraphics[width=0.8\linewidth]{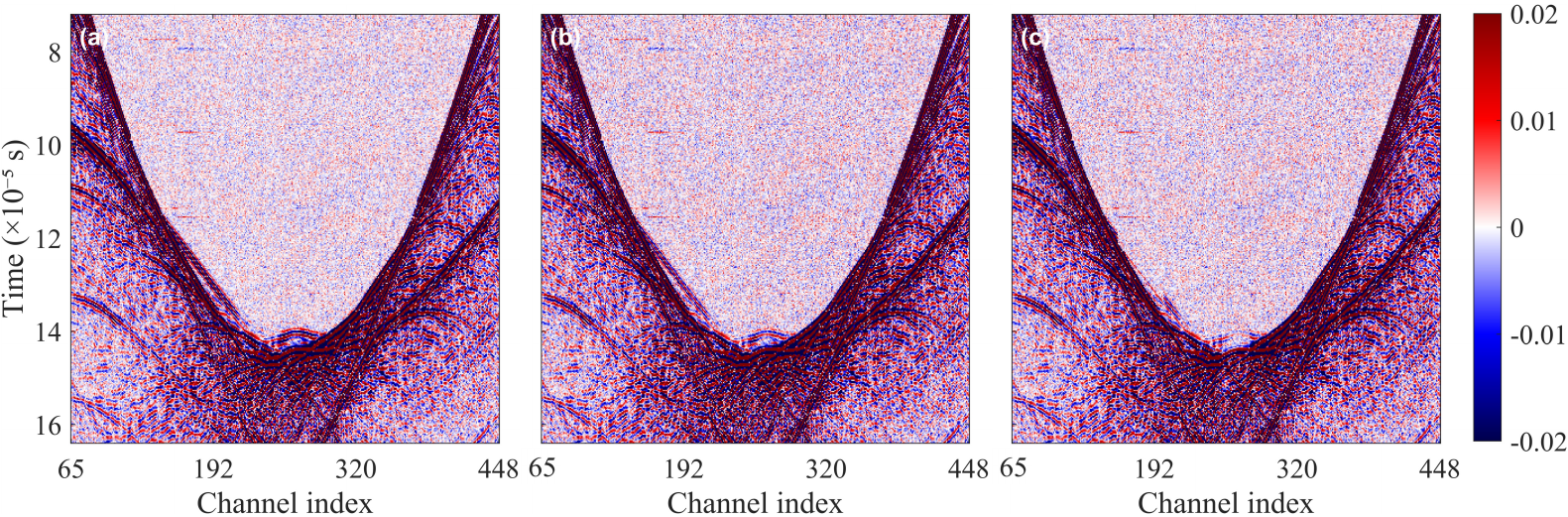}
\caption{Normalized synthetic network inputs at successive pretraining stages. (a) Stage~1 input without additional input perturbation. (b) Stage~2 input with intensity suppression around the annotated first-arrival trajectory. (c) Stage~3 input with domain-randomization perturbations. All images are displayed using the same amplitude window from $-0.02$ to $0.02$.}
\label{fig6}
\end{figure}

\subsection{Evaluation of the Pretrained Model on \textit{In Vitro} Phantom Data}
\label{sub2sec3}
\begin{figure}[htbp]
    \centering
    \includegraphics[width=0.4\linewidth]{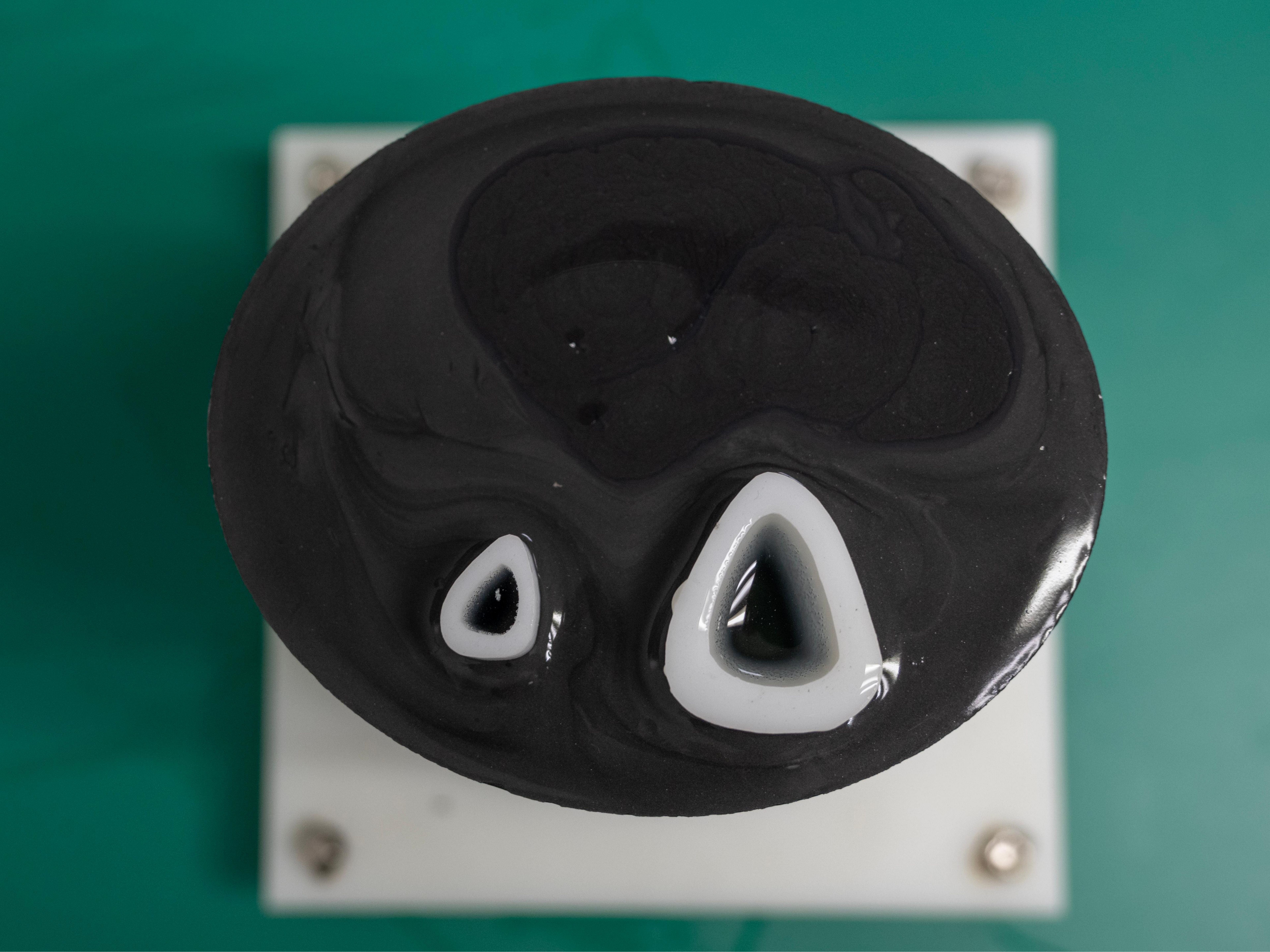}
    \caption{Top-view photograph of the \textit{in vitro} phantom with two hollow bone-mimicking structures.}
    \label{fig7}
\end{figure}

The first-arrival extraction performance of the pretrained network is evaluated using \textit{in vitro} phantom data. The phantom is constructed from human tibia cross-sections extracted from CT images. Two hollow bone-mimicking structures are fabricated by three-dimensional printing using resin with a sound speed of approximately 2600~m/s. These structures are embedded in a tissue-mimicking medium with a sound speed of 1540~m/s and poured into an elliptical mold to break symmetry. Glycerin is injected into the hollow cavities to mimic bone marrow. A top-view photograph showing the cross-sectional configuration of the phantom is presented in Fig.~\ref{fig7}.

Compared with real musculoskeletal tissues, the resin material exhibits lower acoustic-impedance contrast and simpler internal geometry, resulting in reduced scattering and attenuation during wave propagation. This naturally leads to higher SNR in the first-arrival regions, making the extraction task easier than in real scenarios. Moreover, repositioning the phantom within the array produces highly consistent waveform features across acquisitions, which may increase the risk of overfitting if fine-tuning is performed on such limited data variability. Based on these considerations, the pretrained model obtained from the three-stage synthetic training pipeline is applied directly, without additional fine-tuning on the phantom data.

Figure~8(a) shows the network input for one selected transmitter, comprising 384 reordered receiver channels. To facilitate visualization of the wavefront structures in the first-arrival region, the normalized signal amplitudes are clipped to the range from $-0.01$ to 0.01. The same clipping range is used in all subsequent visualizations for consistency. The corresponding binary output from the segmentation network is shown in Fig.~8(b). Compared with the standard synthetic inputs used during training, such as Fig.2(a), the experimental waveform in Fig.~8(a) exhibits more distinct scattered-wave features, which also lead to visible differences in the resulting segmentation mask relative to Fig.~2(b). Nevertheless, the network identifies a plausible first-arrival trajectory. For evaluation, the first foreground pixel in each channel is extracted and overlaid as a red curve in Fig.~8(a).

For comparison, a conventional STA/LTA algorithm is applied independently to each trace. The blue curve in Fig.~8(a) shows the result obtained directly from the raw waveform, which does not provide meaningful first-arrival picks. To improve sensitivity to weak signals, the waveform amplitudes are clipped to $\pm 50$ before STA/LTA analysis, yielding the yellow curve. Although this adjustment improves the result to some extent, the extracted first-arrival times remain noisy and unstable. In contrast, the trajectory identified by the network follows the observed wavefront more consistently and exhibits better continuity than the STA/LTA-based picks.

\begin{figure}[htbp]
    \centering
    \includegraphics[width=\linewidth]{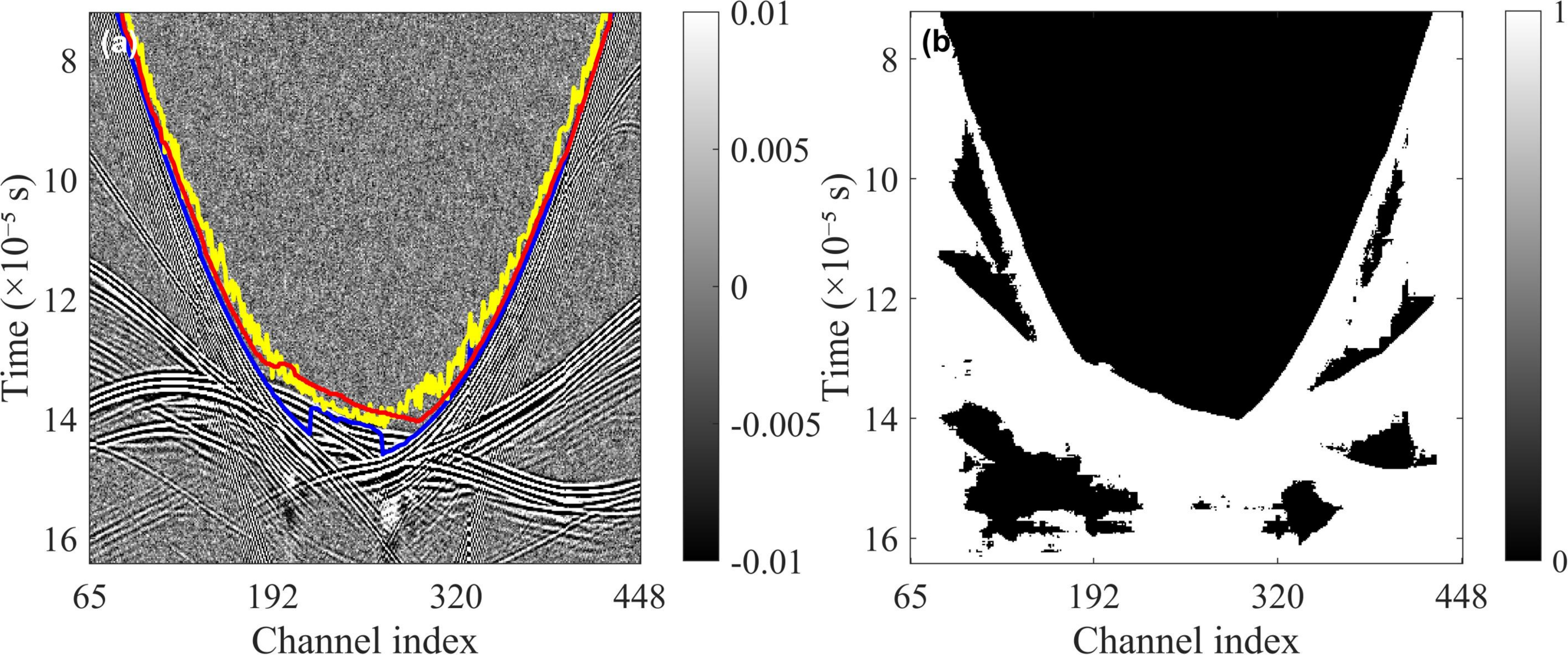}
    \caption{(a) Normalized input waveform data displayed with an amplitude range from $-0.01$ to 0.01, overlaid with first-arrival trajectories obtained by STA/LTA on the raw data (blue), STA/LTA after amplitude clipping to $\pm 50$ (yellow), and the proposed network (red). (b) Binary segmentation mask produced by the network.}
    \label{fig8}
\end{figure}

To assess the impact of first-arrival quality on initial-model construction, HFWI is performed using first-arrival times derived from the STA/LTA method (yellow curve) and from the proposed network. The hybrid inversion minimizes the objective in Eq.~\ref{eq:hfwi_loss} over five frequencies from 0.25 to 0.45~MHz in steps of 0.05~MHz. At each frequency, the model is updated once using steepest descent and then passed to the next frequency. The trade-off weight $\alpha$ is scheduled as {0.75, 0.6, 0.45, 0.3, 0.15}, gradually shifting the emphasis from traveltime fitting to waveform fitting.

The resulting initial models are shown in Fig.~9(a) and Fig.~9(b), constructed using first-arrival times derived from the STA/LTA method and the proposed network, respectively. While both methods recover the overall bone contour, the network-based reconstruction provides a more plausible sound-speed distribution within the cortical-bone regions. In contrast, the STA/LTA-based model underestimates the sound speed, likely because of errors in the extracted first-arrival times.

To further evaluate the effect of initial-model quality, two successive rounds of high-frequency FWI are performed using frequency components from 0.5 to 1.2~MHz with a step size of 0.1~MHz. Each frequency undergoes five conjugate-gradient iterations, and the output of the first round is used as the initialization for the second. During this process, $\alpha$ is fixed at zero, so the objective reduces to conventional FWI, consistent with the final stage of the inversion pipeline described in Section~\ref{sub2sec2}. The final reconstructions are shown in Fig.~9(c) and Fig.~9(d). The model initialized with STA/LTA-derived first-arrival times does not recover the bone morphology as clearly, whereas the model initialized with network-derived first-arrival times yields a more plausible reconstruction. These results illustrate the benefit of improved first-arrival estimates for HFWI-based initialization.

\begin{figure}[h]
    \centering
    \includegraphics[width=0.6\linewidth]{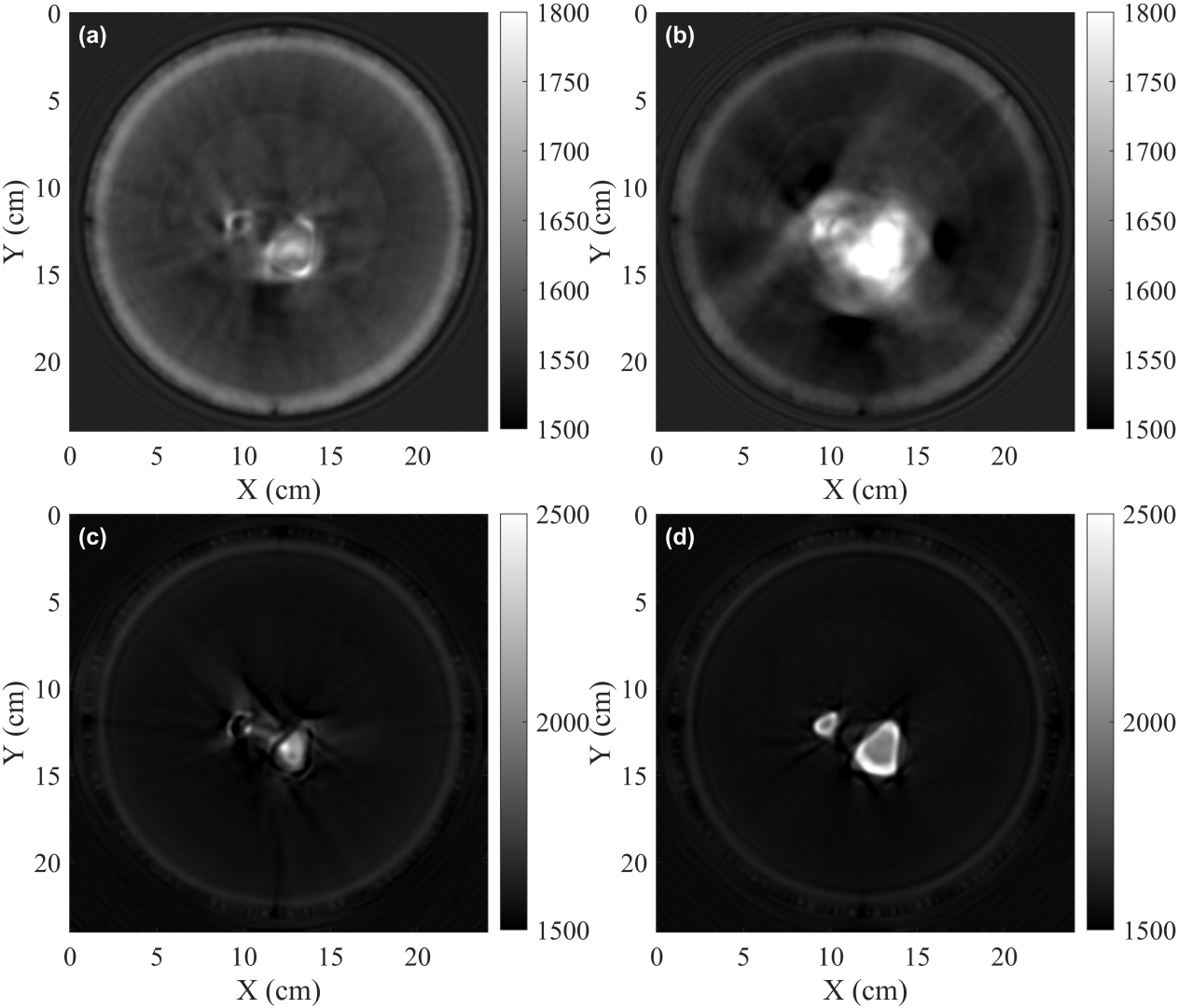}
    \caption{Initial models reconstructed from low-frequency data in the range 0.25--0.45~MHz using first-arrival times derived from (a) STA/LTA and (b) the proposed network. Final reconstructions after subsequent high-frequency FWI using data from 0.5 to 1.2~MHz are shown in (c) and (d), initialized from (a) and (b), respectively.}
    \label{fig9}
\end{figure}

\subsection{Evaluation on Biological Limb Measurements}
\label{sub3sec3}
The network is further evaluated using biological musculoskeletal datasets, including \textit{ex vivo} bovine limbs and \textit{in vivo} human thighs. The evaluation first examines the contribution of the three-stage synthetic pretraining through an ablation study on the \textit{ex vivo} data, and then assesses the effect of decoder-only fine-tuning on both \textit{ex vivo} and \textit{in vivo} measurements. The \textit{ex vivo} dataset is selected for the ablation analysis because its first-arrival references can be annotated more reliably than those of the lower-SNR \textit{in vivo} measurements.

The \textit{ex vivo} experiments are conducted on the bovine-limb cross-section shown in Fig.~1(b). During acquisition, the same settings as those used for synthetic data generation are applied, namely a nominal excitation voltage of 200~Vpp and a receiver gain of 21~dB. Owing to the controllable nature of \textit{ex vivo} measurements, repeated acquisitions with fixed positioning are averaged to improve the SNR in the first-arrival region, thereby facilitating manual annotation.

A total of 256 transmission events are manually annotated from one 512-transmission FMC dataset by selecting every other transmitting element. Among these labeled data, 128 transmissions within the transmitter range from element~\#129 to element~\#384 are used for subsequent decoder-only fine-tuning, while the remaining 128 labeled transmissions are reserved for performance evaluation. This split allows the adapted model to be evaluated on transmitter positions that are not included in experimental fine-tuning.

Before fine-tuning, an ablation study is conducted to evaluate the direct generalization of different synthetic-pretraining configurations to the \textit{ex vivo} measurements. Five configurations are considered: the complete three-stage pretraining strategy, pretraining without Stage~1 warm-up, pretraining without the Stage~2 first-arrival intensity modification, pretraining without the Stage~3 random perturbations, and single-stage training using unmodified simulated data. No experimental data are used to train any of the models in this ablation study. Their performance is evaluated on the same 128 held-out transmissions reserved for the subsequent fine-tuning evaluation. Table~\ref{tab:pretraining_ablation} summarizes the corresponding first-arrival timing and segmentation accuracy.

\begin{table}[h]
\centering
\caption{Ablation study of the stage-wise synthetic pretraining strategy on the \textit{ex vivo} bovine-limb dataset. All reported timing errors are absolute errors, and the mean error therefore corresponds to the mean absolute error (MAE).}
\label{tab:pretraining_ablation}
\vspace{2mm}
\scriptsize

\begin{tabular}{llllll}
\hline
Method & Max error (s) & MAE (s) & Pr (\%) & Re (\%) & F1 (\%) \\
\hline

Full model
& $\mathbf{4.48\times10^{-6}}$
& $1.43\times10^{-6}$
& 97.45
& 99.42
& \textbf{98.42} \\

\hline

w/o Stage 1
& $4.55\times10^{-6}$
& $\mathbf{1.37\times10^{-6}}$
& \textbf{97.60}
& 99.22
& 98.40 \\

\hline

w/o Stage 2
& $5.84\times10^{-6}$
& $1.92\times10^{-6}$
& 96.66
& 99.31
& 97.96 \\

\hline

w/o Stage 3
& $5.20\times10^{-6}$
& $1.48\times10^{-6}$
& 97.57
& 99.07
& 98.31 \\

\hline

Raw simulation
& $6.22\times10^{-6}$
& $1.90\times10^{-6}$
& 96.66
& \textbf{99.52}
& 98.06 \\

\hline
\end{tabular}

\end{table}

As shown in Table~\ref{tab:pretraining_ablation}, the complete three-stage strategy achieves the lowest maximum timing error and the highest F1-score. For the remaining metrics, its performance differs only marginally from the best value in each column. Overall, the complete stage-wise strategy provides the most balanced performance across timing and segmentation metrics.

Following the ablation analysis, the complete three-stage pretrained model is fine-tuned using the designated training subset. Figure~10(a)--(c) shows the results for transmitting element~\#1 from the held-out test set, which is not included in experimental fine-tuning. Figures~10(a) and 10(b) present the segmentation masks before and after fine-tuning, respectively, while Fig.~10(c) compares the corresponding first-arrival trajectories with the STA/LTA result. Compared with the pretrained model, the fine-tuned model produces a visibly more continuous segmentation boundary and a first-arrival trajectory that more closely follows the observed wavefront.

The \textit{in vivo} dataset consists of cross-sectional thigh scans from one male and one female volunteer. Compared with the \textit{ex vivo} measurements, the anatomical differences between the two volunteers and the lower acquisition SNR provide a more challenging test of experimental adaptation. During acquisition, the excitation voltage is reduced to 180~Vpp and the receiver gain to 9~dB to avoid signal saturation and preserve scattered-waveform information for subsequent inversion.

Decoder-only fine-tuning is performed using 256 transmissions from the female-thigh dataset, and the adapted model is then evaluated on the male-thigh dataset. Figures~10(d) and 10(e) show the segmentation outputs before and after fine-tuning, respectively, while Fig.~10(f) compares the corresponding first-arrival trajectories using the same color scheme as in Fig.~10(c). The pretrained model exhibits larger deviations from the observed first-arrival wavefront, whereas the fine-tuned model provides a more coherent trajectory under the lower-SNR \textit{in vivo} condition.

\begin{figure}[h]
\centering
\includegraphics[width=0.8\linewidth]{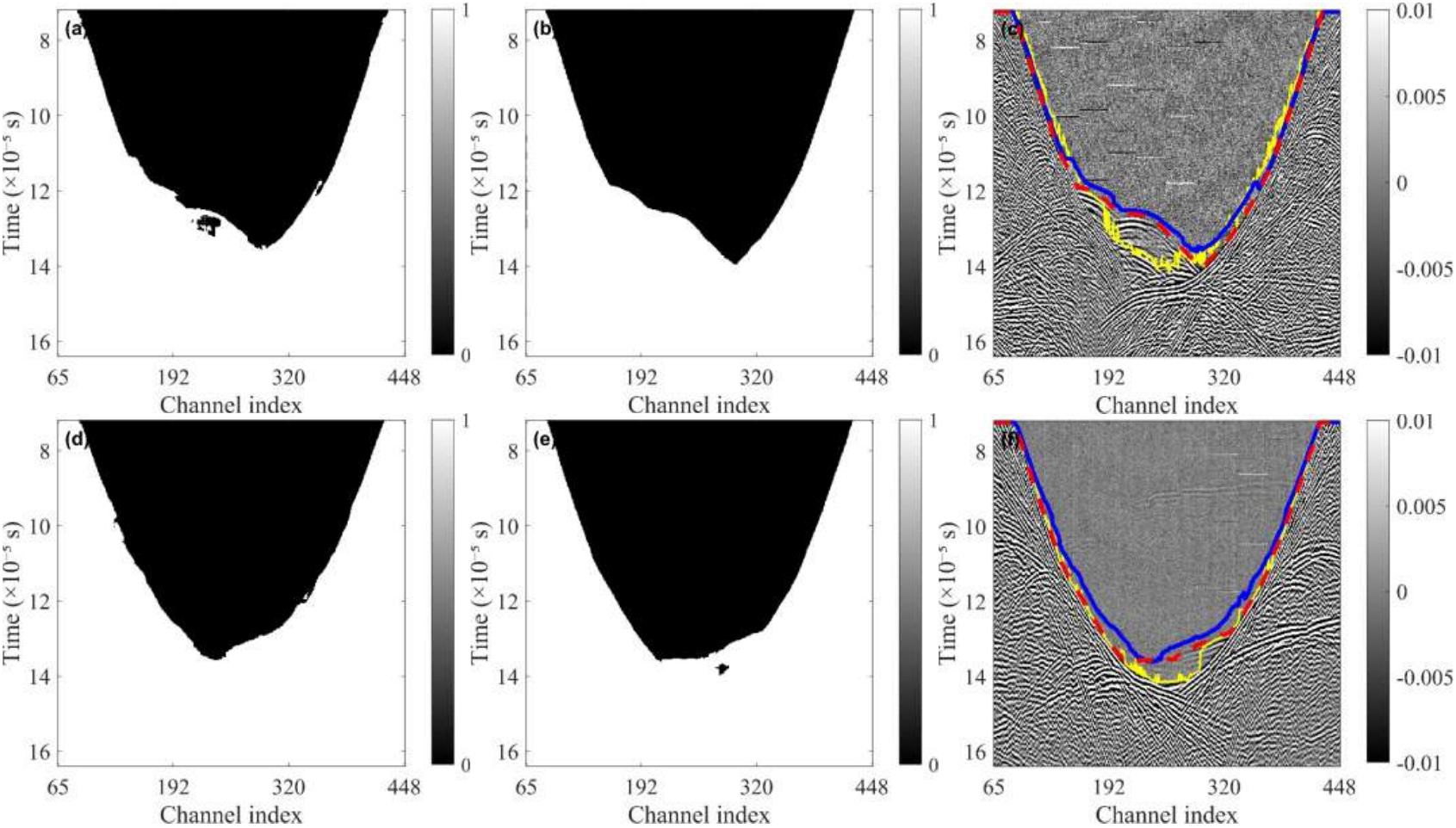}
\caption{Fine-tuning evaluation on biological limb measurements. 
(a), (b) \textit{Ex vivo} bovine-limb data: segmentation masks from the pretrained and fine-tuned models, respectively. 
(c) Corresponding normalized input displayed over an amplitude range of $-0.01$ to 0.01, overlaid with first-arrival trajectories obtained by STA/LTA after amplitude clipping to $\pm 50$ (yellow), the pretrained network (blue), and the fine-tuned network (red). 
(d), (e) \textit{In vivo} human-thigh data: segmentation masks from the pretrained and fine-tuned models, respectively. 
(f) Corresponding normalized input with first-arrival trajectories shown using the same color scheme as in (c).}
\label{fig10}
\end{figure}

The quantitative comparison in Table~\ref{tab:fine_tuning_effect} further quantifies the effect of decoder-only fine-tuning on the same \textit{ex vivo} held-out test set and on the \textit{in vivo} male-thigh data. Fine-tuning reduces the mean absolute first-arrival timing error and improves the F1-score for both datasets. For the \textit{in vivo} case, the increase in precision is accompanied by a slight decrease in recall, while the overall F1-score remains higher after fine-tuning.

\begin{table}[h]
\centering
\caption{Effect of fine-tuning on segmentation-mask accuracy and first-arrival timing error.}
\label{tab:fine_tuning_effect}
\scriptsize
\begin{tabular}{lllllll}
\hline
Dataset & Method & Max error (s) & Mean error (s) & Pr (\%) & Re (\%) & F1 (\%) \\
\hline
\multirow{2}{*}{\shortstack[l]{\textit{Ex vivo}\\bovine limb}}
& Pretrained
& $4.48\times10^{-6}$
& $1.43\times10^{-6}$
& 97.45
& 99.42
& 98.42 \\
\cline{2-7}
& Fine-tuned
& $\mathbf{2.51\times10^{-6}}$
& $\mathbf{4.15\times10^{-7}}$
& \textbf{99.39}
& \textbf{99.50}
& \textbf{99.45} \\
\hline
\multirow{2}{*}{\shortstack[l]{\textit{In vivo}\\human thigh}}
& Pretrained
& $6.55\times10^{-6}$
& $2.61\times10^{-6}$
& 95.62
& \textbf{99.85}
& 97.66 \\
\cline{2-7}
& Fine-tuned
& $\mathbf{2.44\times10^{-6}}$
& $\mathbf{4.61\times10^{-7}}$
& \textbf{99.63}
& 99.63
& \textbf{99.60} \\
\hline
\end{tabular}
\end{table}

Having established the improvement in first-arrival extraction after fine-tuning, the downstream effect of the resulting arrival-time estimates is further evaluated using the male-thigh dataset. At $f=0.25$~MHz, early-stage HFWI models are obtained using $\alpha=0.85$ and five conjugate-gradient iterations under four conditions: without first-arrival guidance, with STA/LTA-derived first-arrival times, with network-derived first-arrival times, and with manually annotated first-arrival times, as shown in Fig.~11(a)--(d), respectively. The first three models are subsequently refined with $\alpha=0$ over the frequency set $\{0.30, 0.35, 0.40, 0.45, 0.50, 0.60, 0.70, 0.80, 0.90, 1.00, 1.10, 1.20\}$~MHz, yielding the reconstructions in Fig.~11(e)--(g). For the manually guided case, the same workflow is followed by two additional FWI cycles to suppress residual artifacts and provide the reference reconstruction in Fig.~11(h), whose structural agreement with co-registered MRI has been reported in \cite{ref41}.

Without first-arrival guidance, the early-stage model in Fig.~11(a) poorly recovers the musculoskeletal sound-speed distribution, and the subsequent reconstruction in Fig.~11(e) contains pronounced streaking artifacts. STA/LTA-derived first-arrival guidance suppresses most of these artifacts and improves the delineation of soft-tissue boundaries; however, residual errors remain along the muscle--fat interface in Fig.~11(f), and the high-sound-speed cortical structure is less clearly recovered. In comparison, the network-guided early-stage model in Fig.~11(c) provides a clearer representation of the cortical structure, and the subsequent reconstruction in Fig.~11(g) exhibits reduced artifacts and improved soft-tissue delineation.

More importantly, the network-guided early-stage model in Fig.~11(c) closely agrees with the manually guided counterpart in Fig.~11(d). After subsequent waveform-based refinement, the network-based reconstruction in Fig.~11(g) also shows close structural consistency with the manually guided reference in Fig.~11(h). These results provide downstream evidence that the residual errors in the network-derived first-arrival estimates remain sufficiently small for the HFWI reconstruction considered in this \textit{in vivo} case.

\begin{figure}[h]
\centering
\includegraphics[width=\linewidth]{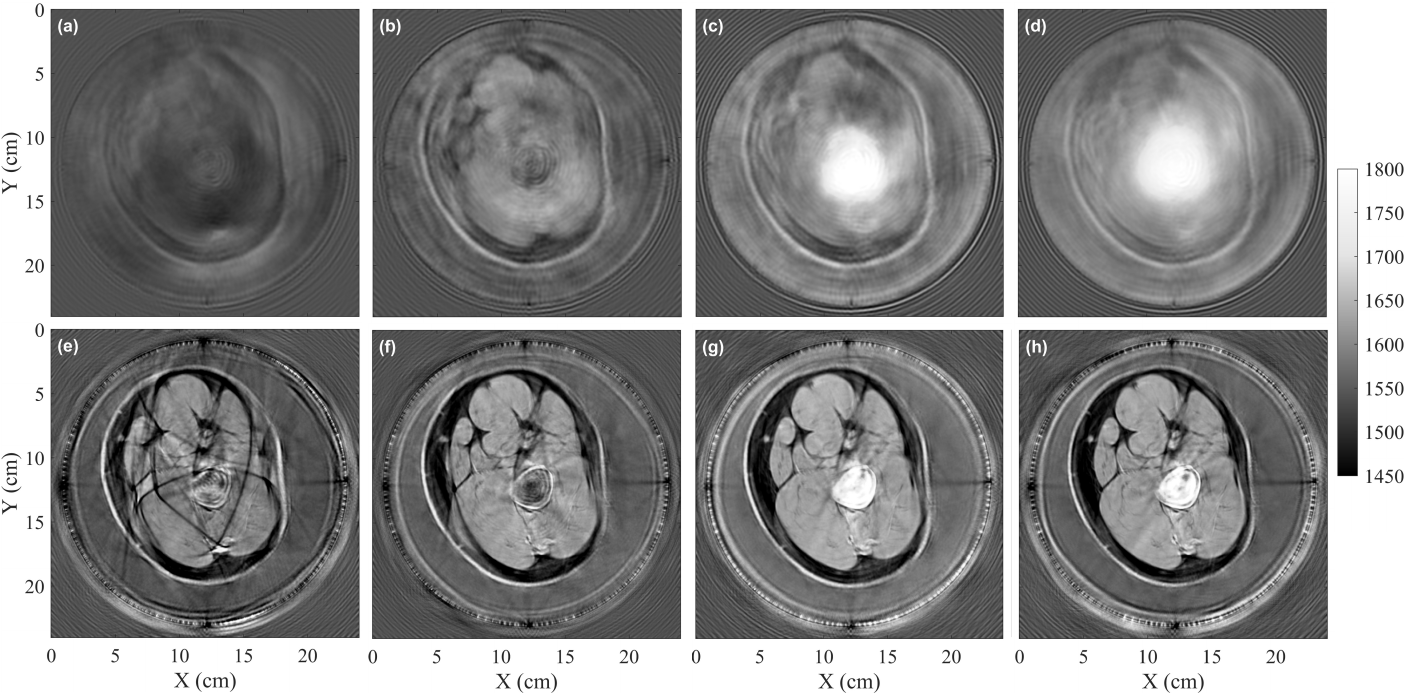}
\caption{Human-thigh reconstructions with different first-arrival inputs. Early-stage HFWI models obtained at 0.25~MHz are shown for (a) no first-arrival guidance, (b) STA/LTA-derived first-arrival times, (c) network-derived first-arrival times, and (d) manually annotated first-arrival times. Corresponding reconstructions after subsequent FWI using data from 0.30 to 1.20~MHz are shown in (e)--(h), respectively. The reconstruction in (h), obtained using manual first-arrival guidance and refined through additional FWI cycles, serves as the reference.}
\label{fig11}
\end{figure}

\section{Discussion}
\label{sec4}
\subsection{Advantages of Segmentation-Based Extraction}
\label{sub1sec4}
\begin{figure}[!htbp]
\centering
\includegraphics[width=0.8\linewidth]{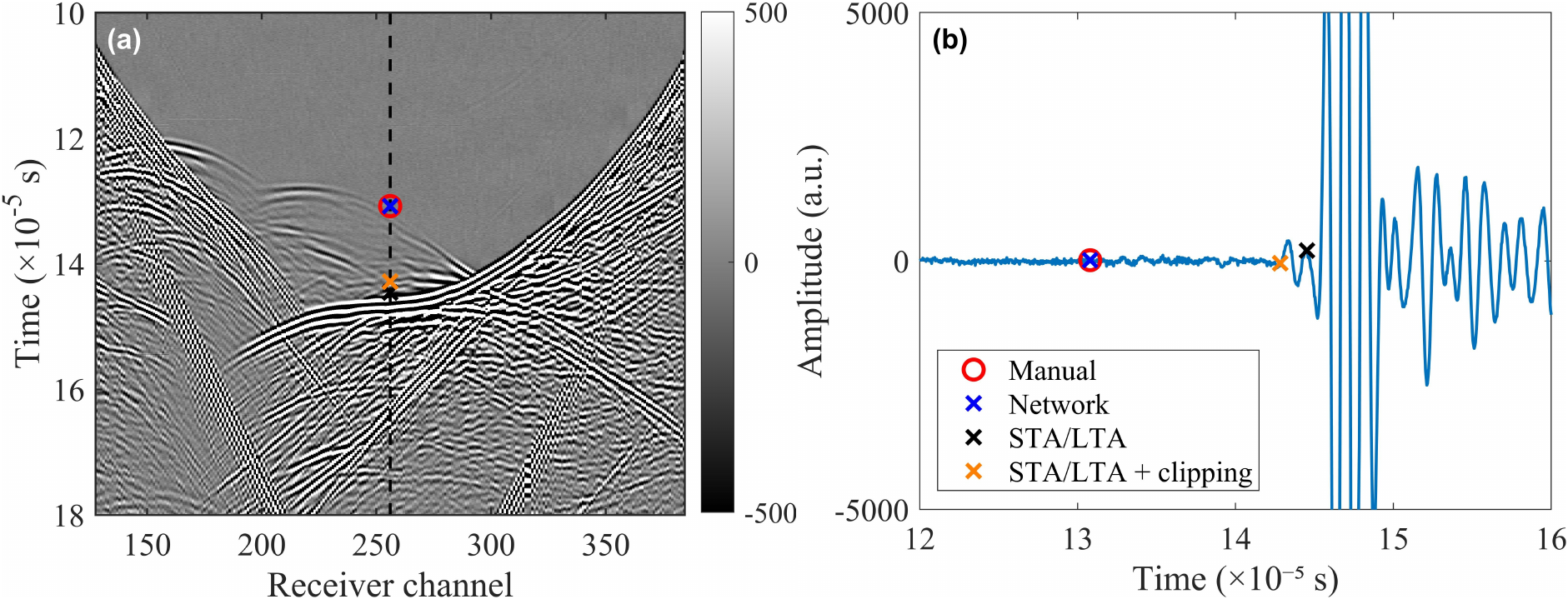}
\caption{First-arrival picks on a single FMC trace from the bovine-limb experiment. The red circle denotes the manual annotation, the black cross denotes STA/LTA applied to the raw signal, the orange cross denotes STA/LTA after amplitude clipping to $\pm 50$, and the blue cross denotes the prediction from the proposed network.}
\label{fig12}
\end{figure}

Conventional trace-wise methods such as STA/LTA identify first arrivals from local energy changes and may become unreliable when the first-arrival response is weak relative to surrounding waveform fluctuations. Figure~\ref{fig12}(a) shows the multichannel waveforms recorded across the retained receiver channels for Src~1, with receiver channel~\#256 indicated by the black dashed line. Figure~\ref{fig12}(b) compares the first-arrival picks on the corresponding single-channel trace. The manually annotated arrival, determined with the aid of repeated-acquisition averaging, lies within a weak-amplitude segment preceding the dominant wave packet, where the estimated local first-arrival SNR of the corresponding raw trace is below $3$~dB. STA/LTA applied to the raw trace is triggered by the later high-amplitude response. Amplitude clipping shifts the estimate toward the manual annotation; however, the corresponding STA/LTA trajectory in Fig.~8(a) still exhibits pronounced trace-to-trace jitter, indicating that preprocessing alone does not ensure spatially consistent picking. By exploiting the two-dimensional continuity of the first-arrival trajectory across receiver channels, the proposed network is less dependent on local amplitude changes and yields a pick closer to the manual annotation.

The quantitative results in Table~\ref{tab:picking_accuracy} further show that the proposed network reduces the mean absolute extraction error on both the \textit{ex vivo} and \textit{in vivo} datasets while shortening the processing time for a complete FMC dataset by more than two orders of magnitude under the implementations used in this study. The lower MAE and substantially shorter runtime support segmentation-based extraction as a more accurate and efficient alternative to conventional STA/LTA picking in the present implementation.

\begin{table}[!htbp]
\centering
\caption{Accuracy and computational efficiency of first-arrival extraction using STA/LTA and the proposed network.}
\label{tab:picking_accuracy}
\small
\resizebox{\linewidth}{!}{%
\begin{tabular}{llllll}
\hline
Dataset & Method & Max error (s) & MAE (s) & Variance (s$^2$) & Time (s) \\
\hline
\multirow{2}{*}{\begin{tabular}[c]{@{}l@{}}\textit{Ex vivo}\\bovine limb\end{tabular}}
& STA/LTA-based
& $1.41\times10^{-4}$
& $1.15\times10^{-4}$
& $3.14\times10^{-10}$
& 480.4201 \\
\cline{2-6}
& Network-based
& $\mathbf{2.51\times10^{-6}}$
& $\mathbf{4.15\times10^{-7}}$
& $\mathbf{1.16\times10^{-12}}$
& \textbf{2.5605} \\
\hline
\multirow{2}{*}{\begin{tabular}[c]{@{}l@{}}\textit{In vivo}\\human thigh\end{tabular}}
& STA/LTA-based
& $1.04\times10^{-5}$
& $2.08\times10^{-6}$
& $6.87\times10^{-12}$
& 507.3462 \\
\cline{2-6}
& Network-based
& $\mathbf{2.44\times10^{-6}}$
& $\mathbf{4.61\times10^{-7}}$
& $\mathbf{2.82\times10^{-12}}$
& \textbf{2.6956} \\
\hline
\end{tabular}%
}
\end{table}

\subsection{Architectural Considerations for First-Break Segmentation}
\label{sub2sec4}

\begin{figure}[h]
\centering
\includegraphics[width=0.6\linewidth]{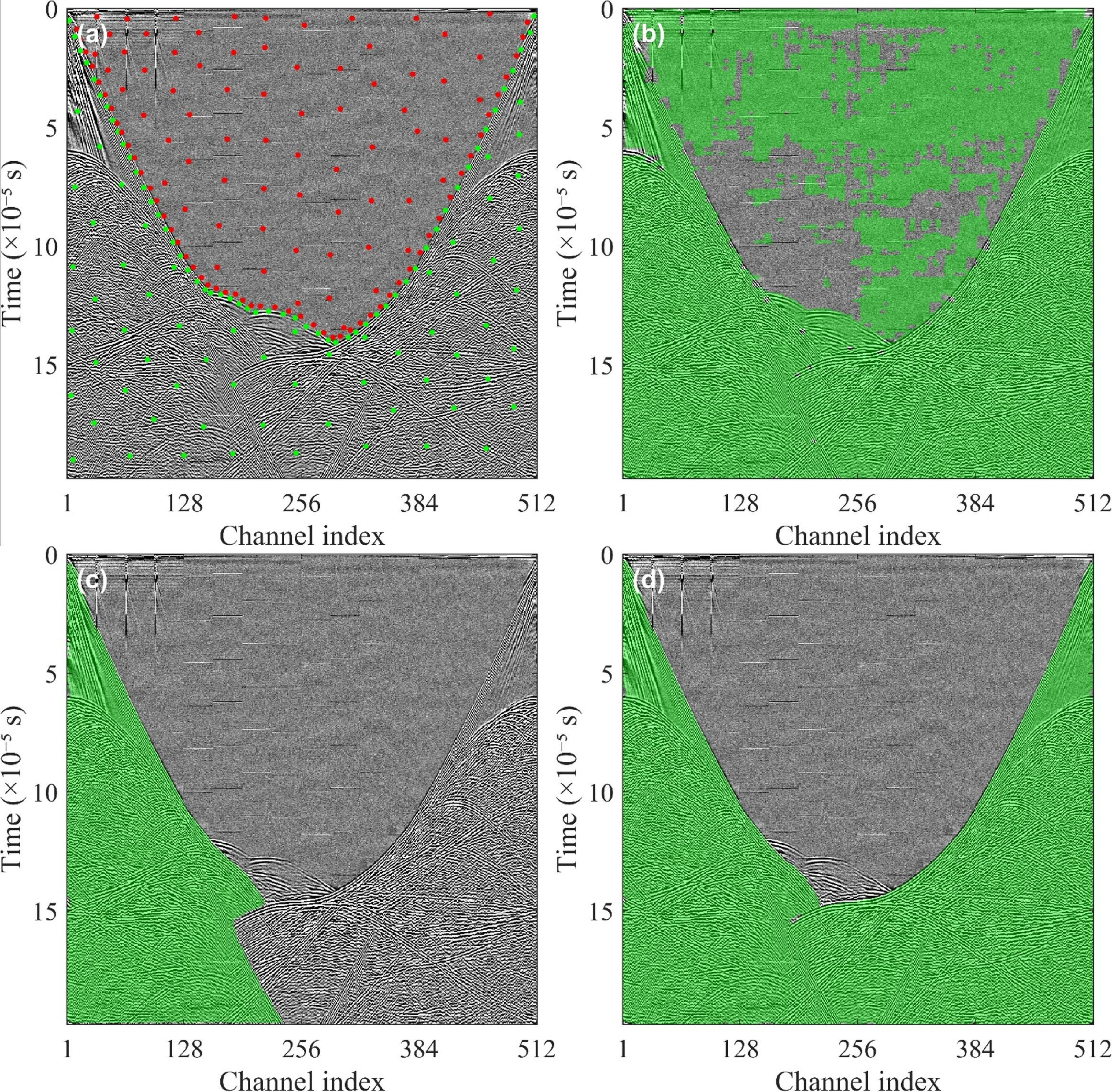}
\caption{Evaluation of a prompt-based Transformer segmentation model on waveform data. (a) Manually placed prompts on the input data, where red points indicate background prompts and green points indicate foreground prompts. (b)--(d) Segmentation masks produced by three SAM variants with increasing parameter counts, shown in the same order.}
\label{fig13}
\end{figure}

First-break segmentation differs from conventional natural-image segmentation because its primary cues are weak localized waveform structures and the spatial continuity of arrival trajectories across neighboring receiver channels, rather than high-level semantic content. To examine whether a general-purpose segmentation model can directly exploit these features, SAM~\cite{ref44} is evaluated without task-specific fine-tuning using dense foreground and background prompts placed with reference to the manually identified boundary. As shown in Fig.~\ref{fig13}, increasing the SAM model size improves the completeness of the predicted acoustic-response region, but even the largest variant fails to recover a reliable first-arrival boundary over the critical receiver-channel range. This result indicates that a Transformer model pretrained on natural images does not directly transfer to the present waveform-segmentation task, even under favorable prompting conditions. It should not be interpreted as a general comparison between CNN and Transformer architectures; however, task-specific training or adaptation of a Transformer would require additional waveform data construction and computational effort, making it a less practical choice under the limited-data and efficiency constraints considered here.

In contrast, the local receptive fields and weight sharing of convolutional networks provide inductive biases that are well matched to the weak but spatially coherent waveform patterns involved in first-arrival extraction. The adopted lightweight U-Net provides a simple multiscale encoder--decoder framework for combining local feature extraction with spatial-continuity modeling, while also supporting the decoder-only adaptation used in this study. Under the limited availability of task-specific data, this design provides sufficient representation capacity while maintaining practical processing speed for the intended workflow, as reflected by the full-FMC processing times in Table~\ref{tab:picking_accuracy} and the downstream HFWI results obtained from the network-derived arrivals. These results indicate that the present architecture provides sufficient extraction quality and practical runtime for the intended inversion workflow; accordingly, rather than increasing network complexity, the present study emphasizes task-specific data construction, stage-wise synthetic pretraining, and limited experimental adaptation to improve transfer from simulated to measured data.

\subsection{Sim2Real Adaptation through Stage-Wise Pretraining and Limited Fine-Tuning}
\label{sub3sec4}
Synthetic simulations provide scalable training data with precise first-arrival labels, but direct transfer to experimental measurements remains limited by the residual Sim2Real gap. The ablation results in Table~\ref{tab:pretraining_ablation} show that the complete three-stage pretraining strategy provides the most balanced performance on the held-out \textit{ex vivo} data. Removing Stage~2 produces the clearest degradation in timing accuracy and F1-score, a trend also observed for the raw-simulation baseline, supporting the benefit of suppressing first-arrival intensity during training to reduce reliance on absolute signal amplitude. Removing Stage~3 also leads to a moderate performance decrease, indicating that random perturbations improve robustness to local waveform discontinuities and ambiguity. By contrast, removing Stage~1 has only a minor effect on the final metrics, suggesting that its primary role is to provide a stable initial learning stage rather than to determine the final cross-domain performance.

Residual discrepancies nevertheless remain after synthetic pretraining, as reflected by the additional improvements obtained through decoder-only fine-tuning in Table~\ref{tab:fine_tuning_effect}. Fine-tuning substantially reduces the mean first-arrival timing error and increases the F1-score for both \textit{ex vivo} and \textit{in vivo} measurements, indicating that a small amount of weakly labeled experimental data is sufficient to correct part of the remaining domain mismatch. Taken together, the results suggest that stage-wise synthetic pretraining provides a robust task-specific initialization, while limited experimental fine-tuning performs the final adaptation to measurement-specific characteristics. This combination reduces dependence on large manually annotated experimental datasets while preserving sufficient first-arrival accuracy for downstream HFWI.

\subsection{Role of First-Arrival Quality in HFWI}
\label{sub4sec4}

In HFWI, first-arrival information is incorporated directly into the early-stage inversion objective as a kinematic constraint, rather than being used to construct a separate initial model through sequential traveltime tomography. This constraint promotes phase alignment between simulated and measured waveforms and thereby reduces the risk of cycle skipping. As the weighting coefficient $\alpha$ decreases, the waveform-fitting term gradually becomes dominant and refines the remaining model details. The effectiveness of this process depends on both the temporal accuracy and the cross-channel consistency of the extracted first-arrival trajectories, because large timing errors or discontinuous picks can weaken the early-stage kinematic guidance.

The main implication of the phantom and \textit{in vivo} results is that the residual errors in the network-derived arrivals remain acceptable for the downstream inversion in this study. The phantom experiment in Fig.~9 shows that the network-derived trajectory provides a more reliable early-stage estimate for subsequent high-frequency FWI than the less consistent STA/LTA result. More importantly, the network-based early-stage model in Fig.~11(c) and final reconstruction in Fig.~11(g) show close visual agreement with those obtained using manually annotated arrivals in Figs.~11(d) and 11(h), respectively. This visual agreement indicates that, within the residual error range observed in this study, exact sample-wise correspondence with manually annotated arrivals is not required to obtain HFWI reconstructions with similar structural characteristics under the tested inversion settings. In contrast, the absence of first-arrival guidance or the use of spatially inconsistent STA/LTA picks leaves more pronounced artifacts. The weaker recovery of cortical structures with STA/LTA is also consistent with its reduced reliability for weak through-bone first arrivals. These results provide downstream validation that the network-derived arrivals can serve as a practical substitute for manually annotated arrivals in the present HFWI workflow.

\section{Conclusion}
\label{sec5}

This study develops a lightweight U-Net-based framework for first-arrival extraction in musculoskeletal USCT, combining stage-wise synthetic pretraining with decoder-only fine-tuning using a limited number of weakly labeled experimental samples. The proposed approach maintains spatially coherent first-arrival extraction under challenging low-SNR conditions, including experimental measurements with local first-arrival SNRs below $3$~dB. Under the implementation used in this study, a complete FMC dataset is processed in 2.56~s for the bovine-limb experiment and 2.69~s for the \textit{in vivo} experiment. Compared with conventional STA/LTA picking, the network yields lower mean absolute timing errors and visually more consistent first-arrival trajectories. When incorporated into HFWI, the network-derived arrivals produce early-stage models and final reconstructions that show close visual agreement with those obtained using manually annotated arrivals, indicating that the observed residual picking errors do not prevent comparable downstream reconstruction behavior under the inversion settings considered in this study.

Several limitations remain. Extremely weak first arrivals can still produce isolated extraction errors, while weak experimental labels may constrain the effectiveness of fine-tuning. In addition, residual Sim2Real mismatch remains because the two-dimensional simulations cannot fully reproduce three-dimensional propagation effects and other experimental complexities. Future work will therefore focus on more diverse experimental data, improved annotation strategies, and incorporating more complete physical modeling to further improve robustness and domain transfer. Network-derived arrival-time information may also be explored as an additional kinematic prior for improving bone-focused musculoskeletal reconstruction.

\section*{Ethics statement}
This study involved human participants. All ethical and experimental procedures were performed in accordance with relevant institutional guidelines and were approved by the Institutional Review Board of Peking University Third Hospital under Approval No.~IRB00006761-M2024690. The participating author provided informed consent for data acquisition and publication. No personally identifiable information is reported in this manuscript.

\section*{Funding}
This work is supported by the National Natural Science Foundation of China (No.~12474461) and the Basic and Frontier Exploration Project Independently Deployed by the Institute of Acoustics, Chinese Academy of Sciences (No.~JCQY202402).

\section*{Data and code availability}
The source code associated with this study is publicly available at
\url{https://github.com/YifeiSUN233/FBseg}.
The repository provides the complete workflows and experimental details for all pretraining procedures, ablation studies, and comparative inference experiments reported in this work.

\bibliographystyle{elsarticle-num}
\bibliography{reference}

@article{ref1,
  title={Imaging techniques for muscle injury in sports medicine and clinical relevance},
  author={Crema, Michel D and Yamada, Andre F and Guermazi, Ali and Roemer, Frank W and Skaf, Abdalla Y},
  journal={Current reviews in musculoskeletal medicine},
  volume={8},
  number={2},
  pages={154--161},
  year={2015},
  publisher={Springer}
}

@article{ref2,
  title={A 3-D ultrasound tomography method for bone morphology evaluation},
  author={Shi, Qinzhen and Zhou, Tianhua and Liu, Yuan and He, Yucheng and Shi, Lingwei and Li, Yifang and Ta, Dean},
  journal={IEEE Transactions on Computational Imaging},
  volume={10},
  pages={17--27},
  year={2024},
  publisher={IEEE}
}

@article{ref3,
  title={Frequency-domain full-waveform inversion-based musculoskeletal ultrasound computed tomography},
  author={Zhou, Chenchen and Xu, Kailiang and Ta, Dean},
  journal={The Journal of the Acoustical Society of America},
  volume={154},
  number={1},
  pages={279--294},
  year={2023},
  publisher={AIP Publishing}
}

@article{ref4,
  title={Quantitative assessment of breast density using transmission ultrasound tomography},
  author={Wiskin, James and Malik, Bilal and Natesan, Rajni and Lenox, Mark},
  journal={Medical physics},
  volume={46},
  number={6},
  pages={2610--2620},
  year={2019},
  publisher={Wiley Online Library}
}

@article{ref5,
  title={Breast density measurements with ultrasound tomography: A comparison with film and digital mammography},
  author={Duric, Neb and Boyd, Norman and Littrup, Peter and Sak, Mark and Myc, Lukasz and Li, Cuiping and West, Erik and Minkin, Sal and Martin, Lisa and Yaffe, Martin and others},
  journal={Medical physics},
  volume={40},
  number={1},
  pages={013501},
  year={2013},
  publisher={Wiley Online Library}
}

@article{ref6,
  title={Frequency domain ultrasound waveform tomography: breast imaging using a ring transducer},
  author={Sandhu, GY and Li, Cuiping and Roy, Olivier and Schmidt, S and Duric, Neb},
  journal={Physics in Medicine \& Biology},
  volume={60},
  number={14},
  pages={5381--5398},
  year={2015},
  publisher={IOP Publishing}
}

@article{ref7,
  title={Clinical Application of Ultrasound Tomography in Diagnosis of Musculoskeletal Diseases.},
  author={Wei, Cong and Zhang, Hui and Ying, Tao and Hu, Bing and Chen, Yini and Li, Hongtao and Zhang, Qiude and Ding, Mingyue and Chen, Jie and Yuchi, Ming and others},
  journal={Advanced Ultrasound in Diagnosis \& Therapy (AUDT)},
  volume={8},
  number={1},
  year={2024}
}

@article{ref8,
  title={The AIUM Practice Parameter for the Performance of the Musculoskeletal Ultrasound Examination},
  author={AIUM and French, Cristy and Hall, Mederic M and ACR and Dahiya, Nirvikar and Allison, Sandra O DeJesus and Levin, Terry and Penna, Rupinder and SPR and Leschied, Jessica and others},
  journal={JOURNAL OF ULTRASOUND IN MEDICINE},
  year={2023},
  publisher={WILEY 111 RIVER ST, HOBOKEN 07030-5774, NJ USA}
}

@article{ref9,
  title={High-resolution imaging without iteration: A fast and robust method for breast ultrasound tomography},
  author={Huthwaite, P and Simonetti, F},
  journal={The Journal of the Acoustical Society of America},
  volume={130},
  number={3},
  pages={1721--1734},
  year={2011},
  publisher={AIP Publishing}
}

@article{ref10,
  title={In vivo breast sound-speed imaging with ultrasound tomography},
  author={Li, Cuiping and Duric, Nebojsa and Littrup, Peter and Huang, Lianjie},
  journal={Ultrasound in medicine \& biology},
  volume={35},
  number={10},
  pages={1615--1628},
  year={2009},
  publisher={Elsevier}
}

@article{ref11,
  title={Quantitative transmission ultrasound tomography: Imaging and performance characteristics},
  author={Malik, Bilal and Terry, Robin and Wiskin, James and Lenox, Mark},
  journal={Medical physics},
  volume={45},
  number={7},
  pages={3063--3075},
  year={2018},
  publisher={Wiley Online Library}
}

@inproceedings{ref12,
  title={Towards ultrasound travel time tomography for quantifying human limb geometry and material properties},
  author={Fincke, Jonathan R and Feigin, Micha and Prieto, Germ{\'a}n A and Zhang, Xiang and Anthony, Brian},
  booktitle={Medical Imaging 2016: Ultrasonic Imaging and Tomography},
  volume={9790},
  pages={470--480},
  year={2016},
  organization={SPIE}
}

@article{ref13,
  title={Ultrasonic computed tomography based on full-waveform inversion for bone quantitative imaging},
  author={Bernard, Simon and Monteiller, Vadim and Komatitsch, Dimitri and Lasaygues, Philippe},
  journal={Physics in Medicine \& Biology},
  volume={62},
  number={17},
  pages={7011--7035},
  year={2017},
  publisher={IOP Publishing}
}

@article{ref14,
  title={Inversion of seismic reflection data in the acoustic approximation},
  author={Tarantola, Albert},
  journal={Geophysics},
  volume={49},
  number={8},
  pages={1259--1266},
  year={1984},
  publisher={Society of Exploration Geophysicists}
}

@article{ref15,
  title={An overview of full-waveform inversion in exploration geophysics},
  author={Virieux, Jean and Operto, St{\'e}phane},
  year={2010}
}

@article{ref16,
  title={Multiscale seismic waveform inversion},
  author={Bunks, Carey and Saleck, Fatimetou M and Zaleski, Stephane and Chavent, Guy},
  journal={Geophysics},
  volume={60},
  number={5},
  pages={1457--1473},
  year={1995},
  publisher={Society of Exploration Geophysicists}
}

@article{ref17,
  title={Building starting models for full waveform inversion from wide-aperture data by stereotomography},
  author={Prieux, Vincent and Lambar{\'e}, Gilles and Operto, St{\'e}phane and Virieux, Jean},
  journal={Geophysical Prospecting},
  volume={61},
  pages={109--137},
  year={2013},
  publisher={Blackwell Publishing Ltd Oxford, UK}
}

@article{ref18,
  title={Quantitative sound speed imaging of cortical bone and soft tissue: Results from observational data sets},
  author={Fincke, Jonathan and Zhang, Xiang and Shin, Bonghun and Ely, Gregory and Anthony, Brian W},
  journal={IEEE Transactions on Medical Imaging},
  volume={41},
  number={3},
  pages={502--514},
  year={2021},
  publisher={IEEE}
}

@article{ref19,
  title={Akaike information criterion statistics},
  author={Sakamoto, Yosiyuki and Ishiguro, Makio and Kitagawa, Genshiro},
  journal={Dordrecht, The Netherlands: D. Reidel},
  volume={81},
  number={10.5555},
  pages={26853},
  year={1986},
  publisher={Taylor \& Francis}
}

@article{ref20,
  title={Automatic P-wave arrival detection and picking with multiscale wavelet analysis for single-component recordings},
  author={Zhang, Haijiang and Thurber, Clifford and Rowe, Charlotte},
  journal={Bulletin of the Seismological Society of America},
  volume={93},
  number={5},
  pages={1904--1912},
  year={2003},
  publisher={Seismological Society of America}
}

@article{ref21,
  title={Fast-AIC method for automatic first arrivals picking of microseismic event with multitrace energy stacking envelope summation},
  author={Long, Yun and Lin, Jun and Li, Bin and Wang, Hongchao and Chen, Zubin},
  journal={IEEE Geoscience and Remote Sensing Letters},
  volume={17},
  number={10},
  pages={1832--1836},
  year={2019},
  publisher={IEEE}
}

@article{ref22,
  title={A comparison of select trigger algorithms for automated global seismic phase and event detection},
  author={Withers, Mitchell and Aster, Richard and Young, Christopher and Beiriger, Judy and Harris, Mark and Moore, Susan and Trujillo, Julian},
  journal={Bulletin of the Seismological Society of America},
  volume={88},
  number={1},
  pages={95--106},
  year={1998},
  publisher={The Seismological Society of America}
}

@article{ref23,
  title={STA/LTA algorithm analysis and improvement of Microseismic signal automatic detection},
  author={Liu, Han and ZHANG, Jian-zhong},
  journal={Progress in Geophysics},
  volume={29},
  number={4},
  pages={1708--1714},
  year={2014},
  publisher={Progress in Geophysics}
}

@article{ref24,
  title={STA/LTA method for picking up the first arrival of natural seismic waves and its improvement analysis},
  author={Qiu, Lei and LI, CaiHua},
  journal={Progress in Geophysics},
  volume={38},
  number={4},
  pages={1497--1506},
  year={2023},
  publisher={Progress in Geophysics}
}

@article{ref25,
  title={Determination of teleseismic relative phase arrival times using multi-channel cross-correlation and least squares},
  author={VanDecar, JC and Crosson, RS},
  journal={Bulletin of the Seismological Society of America},
  volume={80},
  number={1},
  pages={150--169},
  year={1990},
  publisher={The Seismological Society of America}
}

@article{ref26,
  title={Refinement of arrival-time picks using a cross-correlation based workflow},
  author={Akram, Jubran and Eaton, David W},
  journal={Journal of Applied Geophysics},
  volume={135},
  pages={55--66},
  year={2016},
  publisher={Elsevier}
}

@article{ref27,
  title={Seismic waveform classification and first-break picking using convolution neural networks},
  author={Yuan, Sanyi and Liu, Jiwei and Wang, Shangxu and Wang, Tieyi and Shi, Peidong},
  journal={IEEE Geoscience and Remote Sensing Letters},
  volume={15},
  number={2},
  pages={272--276},
  year={2018},
  publisher={IEEE}
}

@article{ref28,
  title={Machine learning for data-driven discovery in solid Earth geoscience},
  author={Bergen, Karianne J and Johnson, Paul A and de Hoop, Maarten V and Beroza, Gregory C},
  journal={Science},
  volume={363},
  number={6433},
  pages={eaau0323},
  year={2019},
  publisher={American Association for the Advancement of Science}
}

@article{ref29,
  title={Applications of deep neural networks in exploration seismology: A technical survey},
  author={Mousavi, S Mostafa and Beroza, Gregory C and Mukerji, Tapan and Rasht-Behesht, Majid},
  journal={Geophysics},
  volume={89},
  number={1},
  pages={WA95--WA115},
  year={2024},
  publisher={Society of Exploration Geophysicists}
}

@article{ref30,
  title={AEnet: Automatic picking of P-wave first arrivals using deep learning},
  author={Guo, Chao and Zhu, Tieyuan and Gao, Yongtao and Wu, Shunchuan and Sun, Jian},
  journal={IEEE Transactions on Geoscience and Remote Sensing},
  volume={59},
  number={6},
  pages={5293--5303},
  year={2020},
  publisher={IEEE}
}

@inproceedings{ref31,
  title={First arrival picking using U-net with Lovasz loss and nearest point picking method},
  author={Yuan, Pengyu and Hu, Wenyi and Wu, Xuqing and Chen, Jiefu and Van Nguyen, Hien},
  booktitle={SEG International Exposition and Annual Meeting},
  pages={D043S138R001},
  year={2019},
  organization={SEG}
}

@incollection{ref32,
  title={Seismic signal augmentation to improve generalization of deep neural networks},
  author={Zhu, Weiqiang and Mousavi, S Mostafa and Beroza, Gregory C},
  booktitle={Advances in geophysics},
  volume={61},
  pages={151--177},
  year={2020},
  publisher={Elsevier}
}

@article{ref33,
  title={First arrival traveltime picking through 3-D U-Net},
  author={Han, Song and Liu, Yujin and Li, Yubing and Luo, Yi},
  journal={IEEE Geoscience and Remote Sensing Letters},
  volume={19},
  pages={1--5},
  year={2021},
  publisher={IEEE}
}

@article{ref34,
  title={A meta-learning-based approach for automatic first-arrival picking},
  author={Li, Hanyang and Sun, Yuhang and Li, Jiahui and Li, Hang and Dong, Hongli},
  journal={IEEE Transactions on Geoscience and Remote Sensing},
  volume={62},
  pages={1--15},
  year={2024},
  publisher={IEEE}
}

@article{ref35,
  title={A comprehensive survey on transfer learning},
  author={Zhuang, Fuzhen and Qi, Zhiyuan and Duan, Keyu and Xi, Dongbo and Zhu, Yongchun and Zhu, Hengshu and Xiong, Hui and He, Qing},
  journal={Proceedings of the IEEE},
  volume={109},
  number={1},
  pages={43--76},
  year={2020},
  publisher={Ieee}
}

@article{ref36,
  title={Using a deep neural network and transfer learning to bridge scales for seismic phase picking},
  author={Chai, Chengping and Maceira, Monica and Santos-Villalobos, Hector J and Venkatakrishnan, Singanallur V and Schoenball, Martin and Zhu, Weiqiang and Beroza, Gregory C and Thurber, Clifford and EGS Collab Team},
  journal={Geophysical Research Letters},
  volume={47},
  number={16},
  pages={e2020GL088651},
  year={2020},
  publisher={Wiley Online Library}
}

@article{ref37,
  title={A little data goes a long way: Automating seismic phase arrival picking at Nabro volcano with transfer learning},
  author={Lapins, Sacha and Goitom, Berhe and Kendall, J-Michael and Werner, Maximilian J and Cashman, Katharine V and Hammond, James OS},
  journal={Journal of Geophysical Research: Solid Earth},
  volume={126},
  number={7},
  pages={e2021JB021910},
  year={2021},
  publisher={Wiley Online Library}
}

@article{ref38,
  title={A deep transfer learning framework for seismic data analysis: A case study on bright spot detection},
  author={El Zini, Julia and Rizk, Yara and Awad, Mariette},
  journal={IEEE Transactions on Geoscience and Remote Sensing},
  volume={58},
  number={5},
  pages={3202--3212},
  year={2019},
  publisher={IEEE}
}

@article{ref39,
  title={Multi-trace joint downhole microseismic phase detection and arrival picking method based on U-Net},
  author={ZHANG, YiLun and YU, ZhiChao and HU, TianYue and HE, Chuan},
  journal={Chinese Journal of Geophysics},
  volume={64},
  number={6},
  pages={2073--2085},
  year={2021},
  publisher={Chinese Journal of Geophysics}
}

@misc{ref41,
      title={Hybrid Full Waveform Inversion Assisted by Rytov Approximation for Musculoskeletal Ultrasound Computed Tomography}, 
      author={Yifei Sun and Yubing Li and Chang Su and Lekang Jiang and Xiangwei Lu and Ligang Cui and He Sun and Weijun Lin},
      year={2026},
      eprint={2605.25139},
      archivePrefix={arXiv},
      primaryClass={physics.med-ph},
      url={https://arxiv.org/abs/2605.25139}, 
}

@misc{ref42,
title={OpenWaves: A Large-Scale Anatomically Realistic Ultrasound-{CT} Dataset for Benchmarking Neural Wave Equation Solvers},
author={Zhijun Zeng and Youjia Zheng and Hao Hu and Zeyuan Dong and Yihang Zheng and Xinliang Liu and Jinzhuo Wang and Zuoqiang Shi and Linfeng Zhang and Yubing Li and He Sun},
year={2025},
url={https://openreview.net/forum?id=u14Y236LwX}
}

@article{ref43,
    author = {Aubry, J.-F. and Tanter, M. and Pernot, M. and Thomas, J.-L. and Fink, M.},
    title = {Experimental demonstration of noninvasive transskull adaptive focusing based on prior computed tomography scans},
    journal = {The Journal of the Acoustical Society of America},
    volume = {113},
    number = {1},
    pages = {84-93},
    year = {2003},
    month = {01},
    issn = {0001-4966},
    doi = {10.1121/1.1529663},
}

@inproceedings{ref44,
  title={Segment anything},
  author={Kirillov, Alexander and Mintun, Eric and Ravi, Nikhila and Mao, Hanzi and Rolland, Chloe and Gustafson, Laura and Xiao, Tete and Whitehead, Spencer and Berg, Alexander C and Lo, Wan-Yen and others},
  booktitle={Proceedings of the IEEE/CVF international conference on computer vision},
  pages={4015--4026},
  year={2023}
}

@article{refSim2Real1,
  title={Generative neural physics enables quantitative volumetric ultrasound of tissue mechanics},
  author={Zeng, Zhijun and Zheng, Youjia and Su, Chang and Wu, Qianhang and Hu, Hao and Dong, Zeyuan and Gao, Shan and Lv, Yang and Tang, Rui and Cui, Ligang and others},
  journal={IEEE Transactions on Pattern Analysis and Machine Intelligence},
  year={2025}
}

@misc{refSim2Real2,
      title={DA-UCT: Self-Supervised Domain-Adaptive Ultrasound Computed Tomography for Rapid Musculoskeletal Sound Speed Reconstruction}, 
      author={Tianyu Liu and Heyu Ma and Aiduo Wang and Peiwen Li and Boyi Li and Ying Li and Dan Li and Chengcheng Liu and Dean Ta},
      year={2026},
      eprint={2605.25024},
      archivePrefix={arXiv},
      primaryClass={cs.CV},
      url={https://arxiv.org/abs/2605.25024}, 
}

@article{Misfit1,
    author = {Métivier, Ludovic and Allain, Aude and Brossier, Romain and Mérigot, Quentin and Oudet, Edouard and Virieux, Jean},
    title = {Optimal transport for mitigating cycle skipping in full-waveform inversion: A graph-space transform approach},
    journal = {Geophysics},
    volume = {83},
    number = {5},
    pages = {R515-R540},
    year = {2018},
    month = {09},
    issn = {0016-8033},
}

\end{document}